\documentclass{article}

\usepackage{microtype}
\usepackage{graphicx}
\usepackage{subcaption}
\usepackage{booktabs} 

\usepackage{hyperref}

\usepackage[preprint]{main}

\usepackage{amsmath}
\usepackage{amssymb}
\usepackage{mathtools}
\usepackage{amsthm}
\usepackage{arydshln}
\usepackage{graphicx}
\usepackage[capitalize,noabbrev]{cleveref}

\usepackage{booktabs}
\usepackage{multirow}
\usepackage{tcolorbox}
\tcbuselibrary{breakable}
\usepackage{ragged2e}
\usepackage{tcolorbox}
\usepackage{xcolor}
\usepackage[T1]{fontenc}

\theoremstyle{plain}

\theoremstyle{definition}

\theoremstyle{remark}

\usepackage[textsize=tiny]{todonotes}

\icmltitlerunning{PaperX: A Unified Framework for Multimodal Academic Presentation Generation with Scholar DAG}

\begin{document}

\twocolumn[
  \icmltitle{PaperX: A Unified Framework for Multimodal Academic Presentation Generation with Scholar DAG}





\newcommand{\affilcasia}{$^1$}
\newcommand{\affilucas}{$^2$}
\newcommand{\affilpku}{$^3$}
\newcommand{\affilthu}{$^4$}
\newcommand{\affilhkustgz}{$^5$}
\newcommand{\caffilucas}{$^{,2}$}
\newcommand{\caffilhkustgz}{$^{,5}$}
\newcommand{\equalmark}{$^*$}           
\newcommand{\leadermark}{$^\ddagger$}   
\newcommand{\corrmark}{$^\dagger$}      

\begin{center}
{\bf Tao Yu\affilcasia\caffilucas\equalmark} \quad
{\bf Minghui Zhang\affilcasia\equalmark\leadermark} \quad
{\bf Zhiqing Cui\affilcasia\caffilhkustgz\equalmark} \quad
{\bf Hao Wang\affilcasia\leadermark} \quad
{\bf Zhongtian Luo\affilcasia\leadermark} \quad
{\bf Shenghua Chai\affilcasia\leadermark}

\vspace{0.1in}

{\bf Junhao Gong\affilpku} \quad
{\bf Yuzhao Peng\affilthu} \quad
{\bf Yuxuan Zhou\affilthu} \quad
{\bf Yujia Yang\affilucas} \quad
{\bf Zhenghao Zhang\affilucas} \quad
{\bf Haopeng Jin\affilcasia\leadermark}

\vspace{0.1in}

{\bf Xinming Wang\affilucas} \quad
{\bf Yufei Xiong\affilcasia\leadermark} \quad
{\bf Jiabing Yang\affilcasia\caffilucas} \quad
{\bf Jiahao Yuan\affilhkustgz} \quad
{\bf Hanqing Wang\affilhkustgz}

\vspace{0.1in}

{\bf Hongzhu Yi\corrmark\affilucas} \quad
{\bf Yan Huang\corrmark\affilcasia\caffilucas} \quad
{\bf Liang Wang\affilcasia\caffilucas}

\vspace{0.15in}

{\small
$^1$CASIA \quad
$^2$UCAS \quad
$^3$Peking University \quad
$^4$Tsinghua University \quad
$^5$HKUST(GZ)
}

\vspace{0.08in}

{\small
$^*$Equal contribution \quad
$^\ddagger$Work done during an internship at CASIA. \quad
$^\dagger$Corresponding author
}

\vspace{0.05in}

{\small
\texttt{yutao2025@ia.ac.cn}, \texttt{zhiqingcui@hkust-gz.edu.cn}
}

\vspace{0.05in}

{
\url{https://github.com/yutao1024/PaperX}
}

\end{center}




  \icmlkeywords{Machine Learning, ICML}

  \vskip 0.3in
]


\makeatletter
\global\icml@noticeprintedtrue  
\@copyrightspace  
\makeatother

\begin{abstract}

Transforming scientific papers into multimodal presentation content is essential for research dissemination but remains labor intensive. Existing automated solutions typically treat each format as an isolated downstream task, leading to redundant processing and semantic inconsistency. We introduce PaperX, a unified framework that models academic presentation generation as a structural transformation and rendering process. Central to our approach is the Scholar DAG, an intermediate representation that decouples the paper's logical structure from its final presentation syntax. By applying adaptive graph traversal strategies, PaperX generates diverse, high quality outputs from a single source. Comprehensive evaluations demonstrate that our framework achieves the state of the art performance in content fidelity and aesthetic quality while significantly improving cost efficiency compared to specialized single task agents.

\end{abstract}

\section{Introduction}

In the rapidly evolving landscape of scientific communication, the dissemination of research findings has transcended the traditional boundaries of static PDF manuscripts, with growing emphasis on \textbf{PPTs} for oral presentations at conferences, \textbf{Posters} for visual interaction during sessions, and \textbf{Promotion (PR)} content for broader social media engagement and accessibility. However, manually crafting these diverse formats is a labor intensive process~\cite{wang2023code4struct, li2025structured}. The recent advance in Generative AI has accelerated the development of specialized agents tailored to automate these individual tasks. Significant progress has been made in specific modalities: for PPT generation, some works\cite{liang2025slidegen, zheng2025pptagent} have demonstrated strong content quality. In the realm of poster generation, various frameworks\cite{pang2025paper2poster, choi2025posterforest, zhang2025postergen} have emerged to tackle the challenge of compressing long documents into single page layouts. Beyond these traditional academic formats, specialized paper agents have also been developed for web adaptation \cite{chen2025paper2web}, video synthesis \cite{zhu2025paper2video}, and PR generation \cite{chen2025autoprletsautomateacademic}.

\begin{figure*}[t]
    \vskip 0.1in
    \centering
    \includegraphics[width=\textwidth]{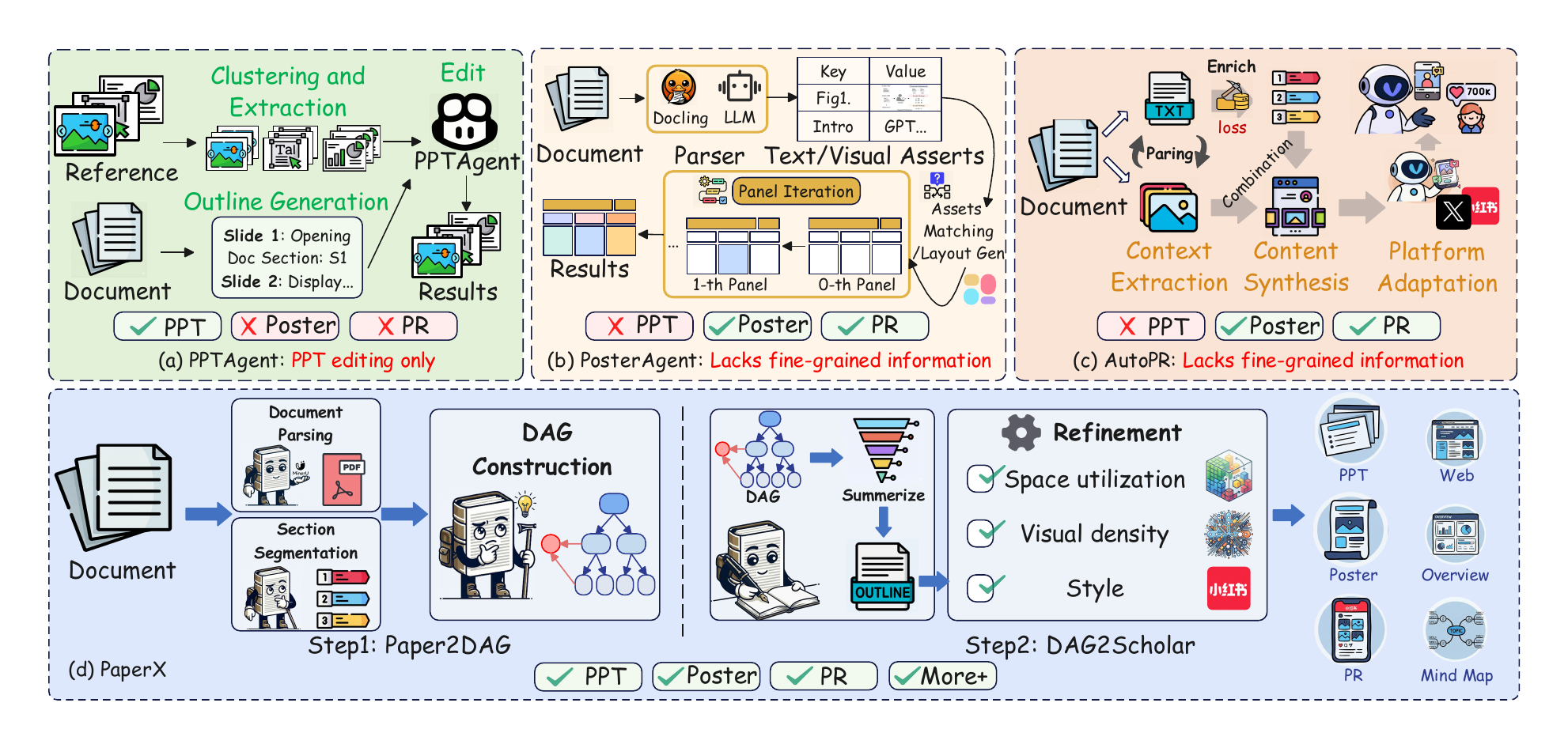}
    \caption{Compared with existing methods, PaperX is able to incorporate and present substantially richer academic content.}
    \label{fig:ablation_study_on_pr_refinement}
\end{figure*}

While these task oriented solutions have achieved success , they expose a fundamental inefficiency: they treat each output format as a distinct downstream task requiring a unique, unrelated pipeline. This leads to redundant semantic processing and inconsistent information presentation, e.g., the key contributions highlighted in a generated poster might differ from those in the accompanying PPTs or PR posts. Furthermore, utilizing separate agents for each format significantly inflates the computational and financial costs of multi-format publishing \cite{ma2025human}.

We argue that existing approaches ignore to take advantage of the intrinsic unity of these tasks. From the perspective of content organization, the differences among presentation formats are primarily manifested in information granularity and structural arrangement, rather than in the underlying semantic content. Thus, the transformation from a paper to different presentation formats can therefore be formalized as a process of multi level content decomposition, aggregation, and reorganization, while preserving semantic consistency. For instance, PPT generation typically requires abstracting detailed content into hierarchical bullet points; poster generation demands selective content pruning and spatial layout under strict space constraints; and PR content emphasizes structured reorganization tailored to dissemination scenarios. Intuitively, an effective intermediate representation should robustly support these operations and satisfy the following requirements: (1) enable hierarchical decomposition and progressive organization to accommodate different presentation granularities; (2) support non-linear content organization and cross region associations to express references, comparisons, and causal dependencies across sections; and (3) maintain structural stability during content pruning and recomposition, ensuring that content selection and reordering do not compromise overall semantic coherence. This further implies that purely sequential representations are insufficient to simultaneously satisfy these requirements, and that a structured representation capable of explicitly modeling hierarchical relationships and cross region dependencies is more appropriate, as shown in Figure \ref{fig:ablation_study_on_pr_refinement}.

Motivated by these insights, we propose PaperX, a structured generation framework that uniformly models multimodal academic content generation as a Paper-to-DAG-to-Scholar process. At the core of PaperX is Scholar DAG (Directed Acyclic Graph), an intermediate representation that parses linear text of a paper into a structured semantic network of arguments, evidence, and figures. Acting analogously to a compiler’s intermediate representation (IR), the Scholar DAG decouples content structure from formatted output rendering, enabling PaperX to generate diverse output formats by selecting information at different granularities from the graph. Specifically, we design distinct generation strategies for three representative modalities: PPTs, Posters, and PRs—and further demonstrate potential extensions of PaperX to highlight its generality and scalability. Our method achieves the best performance on all three benchmarks, PPTEVAL, Paper2Poster, and PRBench.

Our main contributions are summarized as follows:
\begin{itemize}
\item We propose PaperX, a unified framework that models academic presentation generation as a Paper-to-DAG-to-Scholar process, effectively breaking the fragmentation of prior single task agents.
\item We introduce Scholar DAG, an intermediate representation that explicitly models the logical dependencies and hierarchical structure of scientific papers, enabling consistent content distribution across modalities.
\item We design modality specific graph traversal and rendering strategies that achieve state of the-art performance in terms of fidelity, aesthetics, and content quality across key application scenarios.
\end{itemize}

\begin{figure*}[t]
    \vskip 0.1in
    \centering
    \includegraphics[width=\textwidth]{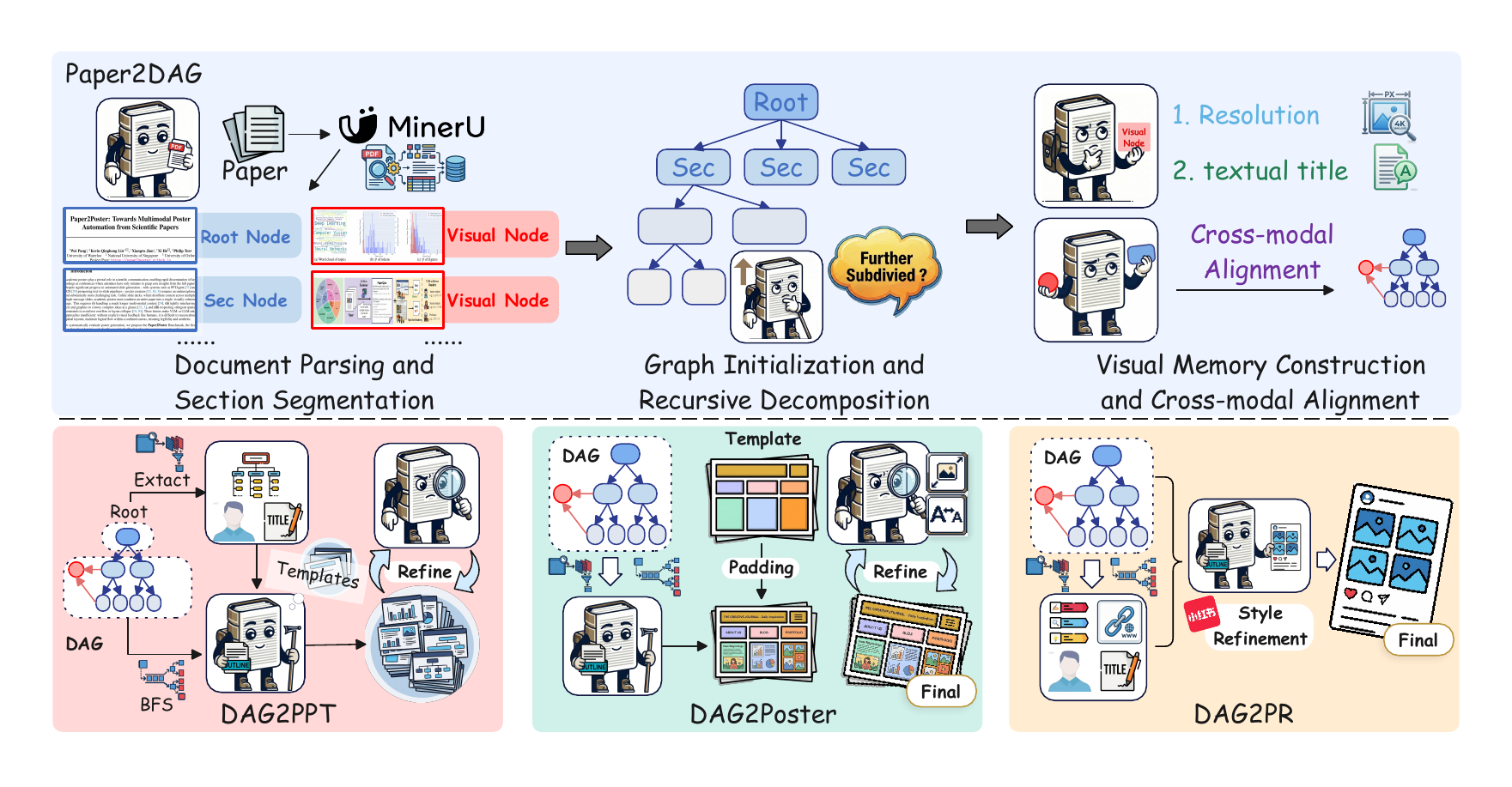}
    \caption{Scholar DAG construction and DAG-driven multimodal academic presentation generation.}
    \label{fig:pipeline}
\end{figure*}

\section{Related Works}

\textbf{Automated Generation of Scientific Presentation} 
Automated scientific media synthesis has evolved from simple text-to-slide extraction \cite{fu2022doc2ppt, wang2023slide4n} toward multimodal narrative construction \cite{harper2024future}. Traditional methods often prioritize single to format outputs like textual outlines or static summaries \cite{shin2024paper}, frequently failing to preserve the semantic fidelity and logical rigor of the original manuscript. Recent research has shifted from template based filling to context aware synthesis, emphasizing the interplay between experimental data, theoretical claims, and visual evidence \cite{tanaka2023slidevqa, goswami2025plotgen}. This evolution, supported by advanced scientific vector graphics and image synthesis \cite{belouadi2023automatikz, zhang2024scimage}, aims to maintain a consistent scientific essence across diverse spatial formats \cite{wang2024scipostlayout}, ensuring that information remains coherent and accurate during the adaptation process.

\textbf{Multimodal Agents for Structured Design} 
Modern design automation has transitioned from rigid layout pipelines \cite{feng2023layoutgpt, lin2025instructlayout} to reasoning centric agentic workflows \cite{li2024survey, xie2024large}. Unlike generative models focused on pixel level synthesis, multimodal agents prioritize the conceptual organization and hierarchical importance of scientific elements \cite{ge2025autopresent, cheng2025graphic}. By leveraging deep reasoning frameworks such as Chain of Thought \cite{wei2022chain, yao2023tree}, these agents interpret complex cross modal relationships, such as aligning methodology descriptions with corresponding graphics \cite{belouadi2025tikzero} to reflect the work's underlying intellectual structure. This paradigm shift, further enabled by efficient UI and interaction modeling \cite{xiao2025efficientuicoder, zhang2025appagent}, positions the agent as a digital curator capable of making strategic decisions on information density and visual emphasis \cite{zheng2025pptagent}, moving beyond simple automation toward professional grade scientific communication.

\section{PaperX}

As shown in Figure \ref{fig:pipeline}, PaperX consists of two stages, Paper2DAG and DAG2Scholar, with DAG2Scholar further divided into DAG2PPT, DAG2Poster, and DAG2PR.

\subsection{Paper2DAG}

\subsubsection{Document Parsing and Section Segmentation}

In this section, we perform structured parsing and section segmentation of the original academic paper before constructing Scholar DAG. The goal of this process is to convert the PDF document into a clean and structured intermediate textual representation, which serves as the foundation for subsequent graph construction and content modeling.

Given a paper in PDF format, denoted by $D$, the input document is mapped to a textual representation $T$ and a set of visual elements $V$. The parsing result is given by
\begin{equation}
(T, V) = \mathcal{P}(D),
\end{equation}
where $\mathcal{P} : D \rightarrow (T, V)$ denotes the document parsing function, and $V = \{v_1, v_2, \ldots, v_m\}$ represents the set of all extracted figures and mathematical formulas from the paper.

Specifically, we adopt the MinerU\cite{niu2025mineru25decoupledvisionlanguagemodel} tool to convert the PDF document into a Markdown representation. During this conversion process, the main body text is mapped into a structured Markdown text $T_{\text{raw}}$, while all figures and formulas are extracted and saved as image files, forming the visual element set $V$. This explicit separation of textual and visual content at the parsing stage facilitates subsequent cross modal structural modeling.

After obtaining the initial Markdown representation $T_{\text{raw}}$, we perform text cleaning and normalization to remove content weakly related to the core content of the paper. This process is formalized as a operator: $T_{\text{clean}} = \mathcal{F}(T_{\text{raw}})$, where $\mathcal{F}(\cdot)$ denotes a filtering function that removes low relevance sections such as \textit{Related Work} and \textit{References}. This step ensures that the resulting textual representation focuses on the main methods, analyses, and conclusions of the paper.

Finally, we segment the cleaned Markdown text $T_{\text{clean}}$ according to the section structure of the paper. Concretely, the text is divided into an ordered set of section textual units:
\begin{equation}
T_{\text{clean}} = \{S_1, S_2, \ldots, S_n\},
\end{equation}
where each $S_i$ corresponds to a section of the original paper.

\subsubsection{Graph Initialization and Recursive Decomposition}

Given the section level textual units $\{S_1, S_2, \ldots, S_n\}$ obtained in the previous step, we construct a hierarchical textual graph and refine it via recursive decomposition. The goal is to organize paper content into a structured directed representation that supports downstream content selection and reorganization while preserving semantic coherence.

\textbf{Graph Initialization} We initialize a textual graph as $G_T=(V_T,E_T)$. We first create a global root node $r$ that encodes the paper metadata, including the title, authors, affiliations, and associated GitHub repository. Each section unit $S_i$ is then mapped to a section node $u_i$:
\begin{equation}
V_T \leftarrow V_T \cup \{r\} \cup \{u_i\}_{i=1}^{n},  \;
E_T \leftarrow E_T \cup \{(r,u_i)\}_{i=1}^{n}.
\end{equation}
We associate each section node $u_i$ with its corresponding text span via a content function $x(u_i) = S_i$, where $x(\cdot)$ returns the textual content stored at a node.

\textbf{Recursive Decomposition via LLM} We employ an LLM driven decomposition operator to iteratively refine each section node into finer-grained semantic units. Let $\mathcal{D}_{\theta}$ denote the decomposition function parameterized by Gemini 3 Pro \cite{comanici2025gemini}, which maps a text span to either a set of sub-spans or an empty set, and the prompt is in Appendix \ref{0}:
\begin{equation}
\mathcal{D}_{\theta}(x) =
\begin{cases}
\{x_1, \ldots, x_k\}, & \text{if } x \text{ is decomposable},\\
\emptyset, & \text{otherwise}.
\end{cases}
\end{equation}
For a node $v$ at depth $d(v)$ with content $x(v)$, we query the LLM and obtain $\mathcal{D}_{\theta}(x(v))$. If $\mathcal{D}_{\theta}(x(v))=\{x_1,\ldots,x_k\}$, we create $k$ child nodes $\{c_j\}_{j=1}^{k}$ such that
\begin{equation}
x(c_j)=x_j,\; V_T \leftarrow V_T \cup \{c_j\}_{j=1}^{k},\;
E_T \leftarrow E_T \cup \{(v,c_j)\}_{j=1}^{k}.
\end{equation}

We recursively apply the same procedure to newly created child nodes until one of the following stopping criteria is met: $\mathcal{D}_{\theta}(x(v))=\emptyset \ \text{or} \ d(v) \geq L$, where $L$ is the maximum recursion depth. After decomposing all section nodes, we obtain the final textual graph $G_T=(V_T,E_T)$, which is rooted at the metadata node $r$ and expands into section level nodes and multi-level semantic sub-nodes.

\subsubsection{Visual Memory Construction and Cross-modal Alignment}

After constructing the textual structure graph $G_T=(V_T,E_T)$, we further incorporate visual elements, including figures and mathematical formulas, into a unified graph representation and establish cross modal alignment between textual and visual content. The objective of this step is to explicitly model visual information while preserving the structural organization of the text, enabling downstream content selection and reorganization across modalities.

\textbf{Visual Node Construction} Let $\mathcal{I} = \{I_1, I_2, \ldots, I_m\}$ denote the set of visual elements extracted during document parsing, where each $I_j$ corresponds to an image of a figure or a mathematical formula. For each visual element $I_j$, we create a corresponding visual node $w_j$, forming the visual node set $V_V = \{w_1, w_2, \ldots, w_m\}$. For each visual node $w_j$, we record its basic visual attributes, including image resolution: $\rho(w_j) = (\mathrm{width}(I_j), \mathrm{height}(I_j))$. To enable semantic referencing and cross modal alignment, each visual node is further associated with a textual title. Let $\tau(w_j)$ denote the title of node $w_j$, which is defined as
\begin{equation}
\tau(w_j) =
\begin{cases}
\mathrm{caption}(I_j), & \text{if an explicit caption is available}, \\
\mathcal{G}_{\theta}(I_j), & \text{otherwise},
\end{cases}
\end{equation}
where $\mathrm{caption}(\cdot)$ extracts the original caption from the document, and $\mathcal{G}_{\theta}(\cdot)$ denotes VLM used to generate a concise semantic description for untitled formulas.

\textbf{Cross modal Alignment} After visual nodes are constructed, we align them with nodes in the textual graph. Let $x(v)$ denote the Markdown text associated with a textual node $v \in V_T$. We define a reference detection function $\mathcal{R}(\cdot)$ that identifies the indices of visual elements explicitly referenced in the text (e.g., via figure, table, or equation identifiers): $\mathcal{R}(x(v)) \subseteq \{1,2,\ldots,m\}$. If $j \in \mathcal{R}(x(v))$, we establish an edge between the textual node $v$ and the visual node $w_j$. Let $E_A$ denote the set of cross modal edges: $E_A = \{(v, w_j) \mid v \in V_T,\ j \in \mathcal{R}(x(v))\}$.

Finally, we obtain a unified graph representation that models textual and visual content: $G = (V_T \cup V_V,\ E_T \cup E_A)$. By modeling figures and formulas as visual nodes and aligning them with corresponding textual contexts, the Scholar DAG captures hierarchical textual structure and cross modal semantic associations, providing a consistent structural foundation for multi-format academic content generation.

\subsection{DAG2Scholar}

\subsubsection{DAG2PPT}

In the \textbf{DAG2PPT} module, we generate academic PPTs through a two stage pipeline: content planning and slide generation, followed by iterative refinement.

\textbf{Content Planning and Slide Generation} Given a Scholar DAG with a global root node $r$, we first generate the textual content for the PPT cover and table of contents based on $r$ and its outgoing edges, which encode the paper title, authors, and high level structure. We then perform a breadth first traversal (BFS) over the subtree rooted at $r$ to select content nodes for slide generation under a predefined page budget.

For each selected node $v$, we employ a LLM to generate a slide level \emph{outline}, defined as
\begin{equation}
\mathrm{outline}(v) = \big(t_{PPT}, \mathcal{I}_{PPT}\big),
\end{equation}
where $t_{PPT}$ denotes a concise textual summary suitable for slide presentation, and $\mathcal{I}_{PPT}$ denotes the set of figures and formulas referenced by node $v$.

Based on the length of $t_{PPT}$ and the cardinality of $\mathcal{I}_{PPT}$, we select an appropriate slide layout from a predefined template library inspired by SlideGen. The LLM then assembles the outline into the selected template, producing an initial version of the slides.

\textbf{Iterative Refinement with VLM Feedback} In the refinement stage, we iteratively improve the generated slides using a vision--language feedback loop. Specifically, we render each slide as an image and provide the slide image together with its corresponding outline to VLM, which identifies potential issues in layout, readability and visual organization, and returns actionable suggestions.

These suggestions, together with the original outline, are then fed back into the LLM to revise the slide content and layout. This refinement process is repeated until one of the following stopping criteria is met: $k \geq K \ \text{or} \ \text{the VLM judges the slide as satisfactory}$, where $k$ denotes the current refinement iteration and $K$ is the maximum allowed number of iterations. The prompts used in this subsubsection are provided in Appendix \ref{1}.

\subsubsection{DAG2Poster}

In the \textbf{DAG2Poster} module, we generate academic posters through a two stage pipeline. The first stage performs content planning and initial placement, while the second stage refines the layout to balance readability and visual density under limited canvas space.

\textbf{Content Planning and Initial Placement} Given a Scholar DAG, we extract the paper title and author information from the root node $r$ and place them into the template header. We then perform a breadth first traversal (BFS) over the subtree rooted at the global node to select nodes for generation.

For each selected content node $v$, we generate a poster level summary and select $1\sim2$ most representative figures or formulas associated with the node. We define the resulting poster unit as an outline:
\begin{equation}
\mathrm{outline}(v) = \big(t_{Poster}, \mathcal{I}_{Poster}\big),
\end{equation}

where $t_{Poster}$ denotes a summary suitable for poster presentation, and $\mathcal{I}_{Poster}$ denotes the selected visual elements.

All outlines are then sequentially placed into the template following a left-to-right, top-to-bottom reading order, producing an initial poster layout.

\textbf{Layout Refinement via Decoupled Optimization} To further improve the initial layout, we introduce a refinement stage inspired by the value, capacity trade off in the knapsack problem. We establish a \emph{text first optimization principle}, as textual content carries the primary semantic information in academic posters and thus requires higher readability, including appropriate font size, line height, and spacing.

We model the poster layout as a constrained optimization problem under a fixed canvas area $A$: $\max \ \mathrm{Readability}(f) + \lambda \cdot \mathrm{VisualCoverage}(i)$,
subject to $\mathrm{Area}(f,i) \leq A$, where $f$ denotes font related parameters (e.g., font size, line height, and spacing), $i$ denotes image size parameters, and $\lambda$ controls the trade off between textual readability and visual density.

To solve this problem, we decouple the two dimensional layout optimization into two subproblems. First, we temporarily fix all images to their minimum acceptable sizes and perform a binary search over font size to determine the maximum font configuration that does not cause layout overflow, thereby ensuring optimal readability. Second, after fixing the font parameters, we greedily expand image sizes using the remaining canvas space to maximize visual coverage and enhance overall visual richness. The prompts used in this subsubsection are provided in Appendix \ref{2}.

\subsubsection{DAG2PR}

In the \textbf{DAG2PR} module, we adopt a two stage pipeline to transform the Scholar DAG into academic promotion content suitable for dissemination on research platforms.

\textbf{Content Planning and Initial Generation} Given a Scholar DAG, we extract the paper metadata from the root node $r$, including the title, authors, affiliations, and the GitHub repository link, and populate them into a PR template. We then perform a breadth first traversal (BFS) over the subtree rooted at the global node to select nodes for generation.

For each selected content node $v$, we generate a academic summary suitable for PR-style communication and select $1\sim2$ most representative figures or formulas associated with the node. We define the resulting PR unit as an outline:
\begin{equation}
\mathrm{outline}(v) = \big(t_{PR}, \mathcal{I}_{PR}\big),
\end{equation}
where $t_{PR}$ denotes the PR oriented textual summary, and $\mathcal{I}_{PR}$ denotes the selected set of visual elements.

All outlines are sequentially inserted into the PR template. After assembling the full template content, we further feed it into a large language model to automatically generate a high level PR title and a set of hashtags: $(\hat{h}, \mathcal{H}) = \mathcal{G}_{\text{PR}}\big(\{\mathrm{outline}(v)\}\big)$, where $\hat{h}$ denotes the generated PR headline, and $\mathcal{H}$ denotes the hashtag set used to summarize key contributions and increase content visibility.

\textbf{Style Refinement for Dissemination} In the second stage, we refine the generated PR content to better match dissemination oriented writing styles. Specifically, we apply a tone refinement function to each outline: $t_{PR}^{\prime} = \mathcal{R}_{\text{style}}(t_{PR})$, where $\mathcal{R}_{\text{style}}(\cdot)$ adjusts the tone of the text to be more engaging and accessible, optionally introducing lightweight stylistic elements (e.g., emojis or expressive markers) without compromising academic professionalism. Through this two stage process, DAG2PR produces PR content that preserves factual accuracy and structural coherence while improving readability, engagement, and platform suitability for academic dissemination. The prompts used in this subsubsection are provided in Appendix \ref{3}.

\begin{table*}[t]
  \centering
  \caption{Performance comparison of presentation generation methods under existing automatic metrics and PPTEval.}

  \resizebox{0.72\textwidth}{!}{%
    \begin{tabular}{lcccccccc}
      \toprule
      \multirow{2}{*}[-0.5ex]{\textbf{Model}} & \multicolumn{4}{c}{\textbf{Existing Metrics}} & \multicolumn{4}{c}{\textsc{\textbf{PPTEval}}} \\
      \cmidrule(lr){2-5} \cmidrule(l){6-9}
      & SR(\%) $\uparrow$ & PPL $\downarrow$ & ROUGE-L$\uparrow$  & FID $\downarrow$ & Content $\uparrow$ & Design $\uparrow$ & Coherence $\uparrow$ & \textbf{Avg.} $\uparrow$ \\
      \midrule
      \multicolumn{9}{l}{\textbf{DocPres (rule-based)}} \\
      \hdashline
      \addlinespace[3pt]
      GPT-4o & -- & 76.42 & 13.28 & -- & 2.98 & 2.33 & 3.24 & 2.85 \\
      Qwen2.5 & -- & 100.4 & 13.09 & -- & 2.96 & 2.37 & 3.28 & 2.87 \\
      \midrule
      \multicolumn{9}{l}{\textbf{KCTV (template-based)}} \\
      \hdashline
      \addlinespace[3pt]
      GPT-4o & 80.0 & 68.48 & 10.27 & -- & 2.49 & 2.94 & 3.57 & 3.00 \\
      Qwen2.5 & 88.0 & \textbf{41.41} & \textbf{16.76} & -- & 2.55 & 2.95 & 3.36 & 2.95 \\
      \midrule
      \multicolumn{9}{l}{\textbf{PPTAgent}} \\
      \hdashline
      \addlinespace[3pt]
      {GPT-4o $\&$  GPT-4o} & \textbf{97.8} & 721.54 & 10.17 & 7.48 & 3.25 & 3.24 & \underline{4.39} & 3.62 \\
      \addlinespace[2pt]
      {Qwen2-VL $\&$  Qwen2-VL} & 43.0 & 265.08 & 13.03 & \underline{7.32} & 3.13 & 3.34 & 4.07 & 3.51 \\
      \addlinespace[2pt]
      {Qwen2.5 $\&$  Qwen2-VL} & \underline{95.0} & 496.62 & \underline{14.25} & \textbf{6.20} & 3.28 & 3.27 & \textbf{4.48} & 3.67 \\
      \midrule
      \multicolumn{9}{l}{\textbf{Ours}} \\
      \hdashline
      \addlinespace[3pt]
      \textbf{PaperX(GPT-4o)} & -- & 73.40 & 7.81 & -- & \underline{3.64} & \textbf{4.27} & 3.74 & \underline{3.88} \\
      \textbf{PaperX(Gemini-3-Pro)} & -- & \underline{62.47} & 6.45 & -- & \textbf{3.69} & \underline{4.23} & 3.84 & \textbf{3.92} \\
      \bottomrule
    \end{tabular}
  }
  \label{tab:performance-comparison}
\end{table*}

\begin{table*}[t]
    \centering 
    \caption{Performance comparison of poster generation methods on the Paper2Poster benchmark.}
    \label{poster-table}
    \resizebox{0.9\textwidth}{!}{%
      \begin{tabular}{l ccc cccc cccc c}
        \toprule
        \multirow{3}{*}[-1ex]{\textbf{Model}} & \multicolumn{3}{c}{\textbf{Vis. quality \& Txt. coherence}} & \multicolumn{9}{c}{\textbf{VLM-as-Judge}} \\
        \cmidrule(lr){2-4} \cmidrule(lr){5-13}
        
        & \multirow{2}{*}[-0.5ex]{Vis. Sim. $\uparrow$} & \multirow{2}{*}[-0.5ex]{PPL $\downarrow$} & \multirow{2}{*}[-0.5ex]{Fig. Rel. $\uparrow$} 
        & \multicolumn{4}{c}{\textbf{Aesthetic Score $\uparrow$}} & \multicolumn{4}{c}{\textbf{Information Score $\uparrow$}} & \multirow{2}{*}[-0.5ex]{\textbf{Overall $\uparrow$}} \\
        \cmidrule(lr){5-8} \cmidrule(lr){9-12}
        
        & & & & Element & Layout & Engage & \textbf{Avg.} & Clarity & Content & Logic & \textbf{Avg.} & \\
        \midrule
        
        \multicolumn{13}{l}{\textbf{Oracle Methods}} \\ 
        \hdashline
        \addlinespace[3pt]
        Paper & 0.53 & \textbf{4.60} & 0.22 & \underline{4.05} & 3.89 & 2.80 & 3.58 & 4.00 & \textbf{4.68} & \textbf{3.98} & \textbf{4.22} & \textbf{3.90} \\
        GT Poster & \textcolor{gray}{1.00} & 11.26 & 0.21 & \textbf{4.07} & 3.90 & 2.70 & 3.56 & \textbf{4.09} & \underline{3.96} & \underline{3.89} & \underline{3.98} & 3.77 \\
        \midrule
        
        \multicolumn{13}{l}{\textbf{End-to-End Methods}} \\ 
        \hdashline
        \addlinespace[3pt]
        4o-HTML & 0.52 & 9.86 & – & 3.53 & 3.82 & 2.72 & 3.36 & 3.94 & 3.64 & 3.47 & 3.68 & 3.52 \\
        4o-Image & \textbf{0.76} & 77.13 & 0.21 & 2.93 & 3.02 & 2.75 & 2.90 & 1.05 & 2.04 & 2.22 & 1.77 & 2.33 \\
        \midrule
        
        \multicolumn{13}{l}{\textbf{Multi-Agent Methods}} \\
        \hdashline
        \addlinespace[3pt]
        OWL-4o & 0.54 & 11.46 & – & 2.76 & 3.62 & 2.56 & 2.98 & 3.92 & 2.89 & 3.36 & 3.39 & 3.19 \\
        PPTAgent-4o & 0.50 & 6.20 & 0.16 & 2.49 & 3.05 & 2.45 & 2.66 & 2.05 & 1.26 & 1.38 & 1.56 & 2.11 \\
        \midrule
        
        \multicolumn{13}{l}{\textbf{PosterAgent Variants}} \\
        \hdashline
        \addlinespace[3pt]
        PosterAgent-4o & \underline{0.75} & 8.31 & \underline{0.24} & 3.95 & 3.86 & \underline{2.93} & 3.58 & 4.03 & \underline{3.96} & 3.60 & 3.86 & 3.72 \\
        PosterAgent-Qwen & \underline{0.75} & 8.81 & \underline{0.24} & 3.93 & 3.67 & 2.89 & 3.50 & 3.95 & 3.85 & 3.68 & 3.83 & 3.66 \\
        \midrule

        \multicolumn{13}{l}{\textbf{Ours}} \\
        \hdashline
        \addlinespace[3pt]
        \textbf{PaperX(GPT-4o)} & 0.72 & 5.85 & \textbf{0.25} & 3.95 & \underline{3.94} & \textbf{3.02} & \underline{3.64} & 4.05 & 3.93 & 3.80 & 3.92 & 3.78 \\
        \textbf{PaperX(Gemini-3-Pro)} & 0.72 & \underline{5.69} & \textbf{0.25} & 3.95 & \textbf{3.97} & \textbf{3.02} &\textbf{ 3.66} & \underline{4.07} & 3.95 & 3.82 & 3.95 & \underline{3.80} \\
        \bottomrule
      \end{tabular}%
    }
    \vskip -0.1in
\end{table*}

\begin{table*}[t]
    \centering
    \caption{Performance comparison of PaperX and AutoPR on the PRBench benchmark.}
    \label{tab:pragent-performance}
    \resizebox{0.95\textwidth}{!}{
        \begin{tabular}{lccccccccccccc}
          \toprule
          \multirow{2}{*}[-2ex]{\textbf{Model}} & \multicolumn{2}{c}{\textbf{Fidelity $\uparrow$}} & \multicolumn{6}{c}{\textbf{Engagement $\uparrow$}} & \multicolumn{4}{c}{\textbf{Alignment $\uparrow$}} & \multirow{2}{*}[-2ex]{\textbf{Overall $\uparrow$}} \\
          \cmidrule(lr){2-3} \cmidrule(lr){4-9} \cmidrule(lr){10-13}
            & A\&T & Factual & \multirow{2}{*}{Hook} & Logical & Visual & \multirow{2}{*}{CTA} & Prof. & Broad & Context & Vis-Txt & \multirow{2}{*}{Hashtag} & Plat. & \\
            & Acc. & Score &  & Attr. & Attr. &  & Pref. & Pref. & Rel. & Int. &  & Pref. & \\
            \midrule
            
            \multicolumn{14}{l}{\textbf{PRAgent}} \\
            \hdashline
            \addlinespace[3pt]
            Qwen2.5-VL-7B-Ins & 62.17 & 57.89 & 62.57 & 58.33 & 59.32 & 15.62 & 66.41 & 74.61 & 57.40 & 60.61 & 50.26 & 70.31 & 57.96 \\
            InternVL3-14B  & 64.78 & 55.91 & 75.26 & 67.06 & 73.05 & 52.80 & 73.05 & 92.19 & 80.79 & 71.55 & 53.22 & 87.89 & 70.63 \\
            Qwen2.5-VL-32B-Ins  & 72.85 & \textbf{72.49} & 74.80 & 82.03 & \textbf{75.33} & 51.69 & \underline{98.05} & \textbf{100.00} & 83.82 & 75.03 & \textbf{61.65} & 96.48 & \underline{78.69} \\
            Qwen3-32B\textsubscript{T} & 70.31 & 64.94 & 75.00 & \underline{83.72} & 74.61 & 42.32 &\textbf{ 99.22} & \textbf{100.00} & \underline{86.91} & 75.39 & 60.71 & \textbf{99.22} & 77.70 \\
            Qwen3-235B-A22B\textsubscript{T} & 66.80 & \underline{66.92} & 75.33 & 83.69 & 74.87 & 42.58 & 97.66 & \textbf{100.00} & \textbf{87.17} & 75.10 & \underline{61.13} & 97.66 & 77.41 \\
            GPT-4o & 66.32 & 45.94 & 75.00 & 75.22 & \underline{74.89} & 49.07 & 77.93 & 98.24 & 81.83 & 74.17 & 52.08 & 97.66 & 72.36 \\
            Gemini-2.5-Pro & 71.81 & 63.14 & 74.47 & \textbf{85.97} & 73.89 & 45.44 & 97.27 & \underline{99.22} & 86.04 & 74.58 & 58.40 & \underline{98.05} & 77.36 \\
            \midrule
            
            \multicolumn{14}{l}{\textbf{Ours}} \\
            \hdashline
            \addlinespace[3pt]
            \textbf{PaperX(GPT-4o)} & \underline{79.80} & 56.04 & \textbf{92.51} & 66.24 & 73.18 & \textbf{98.24} & 73.05 & 94.14 & 61.36 & \underline{86.39} & 54.36 & 70.31 & 75.47 \\
            \textbf{PaperX(Gemini-3-Pro)} & \textbf{100.00} & 64.35 & \underline{92.23} & 66.07 & 74.74 & \underline{97.09} & 82.94 & 95.24 & 61.51 & \textbf{87.17} & 60.05 & 70.24 & \textbf{79.27} \\
            \bottomrule
        \end{tabular}
    } 
\end{table*}

\begin{table}[t]
  \centering
  \caption{Performance comparison of poster generation methods on the Paper2Poster benchmark under PaperQuiz Evaluation.}
  \resizebox{0.9\columnwidth}{!}{%
      \begin{tabular}{l c c c} 
        \toprule
        \textbf{Model} & \textbf{Verbatim $\uparrow$} & \textbf{Interpretive $\uparrow$} & \textbf{Overall $\uparrow$} \\
        \midrule
        \multicolumn{4}{l}{\textbf{Oracle Methods}} \\
        \hdashline
        \addlinespace[3pt]
        Paper & \textbf{67.20} & 65.05 & \textbf{66.12} \\ 
        GTPoster & \underline{54.93} & 63.37 & 59.15 \\
        \midrule
        
        \multicolumn{4}{l}{\textbf{End-to-End Methods}} \\
        \hdashline
        \addlinespace[3pt]
        4o-HTML & 50.23 & 62.96 & 56.59 \\ 
        4o-Image & 39.93 & 60.43 & 50.18 \\
        \midrule

        \multicolumn{4}{l}{\textbf{Multi-Agent Methods}} \\
        \hdashline
        \addlinespace[3pt]
        OWL-4o & 39.92 & 62.16 & 51.04 \\ 
        PPTAgent-4o & 25.81 & 36.68 & 31.25 \\
        \midrule
        
        \multicolumn{4}{l}{\textbf{PosterAgent Variants}} \\
        \hdashline
        \addlinespace[3pt]
        PosterAgent-4o   & 51.06 & 65.35 & 58.21 \\
        PosterAgent-Qwen & 50.30 & 64.62 & 57.46 \\
        \midrule
        
        \multicolumn{4}{l}{\textbf{Ours}} \\
        \hdashline
        \addlinespace[3pt]
        \textbf{PaperX(GPT-4o)}     & 44.58 & \underline{73.92} & 59.25 \\
        \textbf{PaperX(Gemini-3-Pro)}       & 45.81 & \textbf{74.37} & \underline{60.09} \\
        \bottomrule
      \end{tabular}
  }
  \label{poster-table-quiz}
  \vskip -0.1in
\end{table}

\section{Experiments}

\subsection{Benchmarks and Settings}

We evaluate the proposed method on three public benchmarks, PPTEVAL, Paper2Poster, and PRBench, which correspond to slide generation, poster generation, and academic PR generation tasks, respectively. For the performance of baseline methods, we directly adopt the original scores reported in the respective benchmarks. Since these results typically correspond to the optimal configurations used by the original authors, this setting reflects the best known performance of existing methods on each task and avoids unfair bias introduced by reimplementation.

Regarding model configuration, we consistently employ Gemini 3 Pro as LLM during the Scholar DAG construction stage to ensure consistency in the structured modeling process. In the downstream content generation stage for PPT, Poster, and PR generation, we use GPT-4o and Gemini 3 Pro, treating them as the primary comparison baselines to assess differences in generation capabilities across different LLMs under the same structured intermediate representation. The hyperparameters are provided in Appendix \ref{4}.

\subsection{Main Results}

\textbf{Stronger content expression, visual presentation, and language fluency.} As shown in Table \ref{tab:performance-comparison}, under two generation model configurations, PaperX (GPT-4o) and PaperX (Gemini-3-Pro) achieve the highest average PPTEval scores respectively. A breakdown of the metrics further shows that our method exhibits particularly strong performance on the Content and Design dimensions, indicating that the generated PPTs are superior in both content completeness and visual presentation quality. Meanwhile, our method maintains stable and competitive performance on Coherence, suggesting a consistent overall narrative structure. In terms of traditional automatic evaluation metrics, our approach achieves substantially lower perplexity (PPL) than PPTAgent, reflecting improved language fluency. Taken together, these results indicate that the structured content modeling based on Scholar DAG facilitates more effective content selection and organization, while the iterative refinement further enhances the overall design quality of the PPTs.

\textbf{Better cross modal relevance, textual coherence, and overall readability.} As shown in Table \ref{poster-table}, under rule based metrics, our method achieves the best performance on Textual Coherence (PPL↓) and Figure Relevance (Fig. Rel.↑): PaperX (Gemini-3-Pro) attains the lowest PPL of 5.69, and both configurations reach the highest Fig. Rel. score of 0.25. These results indicate that the generated posters exhibit more fluent and predictable text, while the selected figures and formulas are more semantically aligned with their surrounding textual context. Under the VLM-as-Judge evaluation, which better reflects human subjective preferences, our method also achieves the best overall performance among non-oracle approaches. In particular, PaperX (Gemini-3-Pro) obtains an Overall score of 3.80, and remains leading or tied for the highest scores in both aesthetic and information related dimensions. Overall, the structured content selection based on Scholar DAG improves information organization, while the layout refinement further enhance visual presentation and overall readability of the generated posters.

\textbf{Stronger transmission of high level semantic understanding.} Under the PaperQuiz evaluation (Table \ref{poster-table-quiz}), our method achieves the best performance on the Interpretive dimension, significantly outperforming existing methods. These results indicate that the structured content selection based on Scholar DAG more effectively highlights key contributions and argumentative structure, thereby improving the poster’s ability to convey high level understanding.

\textbf{Higher overall communication quality and platform adaptability.} As shown in Table \ref{tab:pragent-performance}, our method achieves the best overall performance on PRBench, with PaperX (Gemini-3-Pro) attaining an Overall score of 79.27, outperforming all competing methods. A breakdown of the results shows that our approach demonstrates advantages across all three evaluation dimensions. In terms of Fidelity, we achieve a perfect score of 100.00 on authorship and title accuracy (A\&T Acc.), indicating that key attribution information is consistently and prominently presented. For Engagement, our method attains high scores on metrics such as Hook and CTA, suggesting stronger ability to attract readers and encourage follow up actions. Regarding Alignment, we remain leading or competitive on visual–text integration (Vis-Txt Int.) and platform preference (Plat. Pref.), reflecting better consistency with platform specific styles and dissemination strategies. Overall, these results indicate that the structured content selection based on Scholar DAG improves factual consistency and information focus, while publication oriented tone refinement further enhance platform adaptability and communication effectiveness.

\subsection{Further Analysis}

\subsubsection{Qualitative Analysis}

As shown in Fig.\ref{fig:ablation_study_on_dag}, we compare PPTs generated using Scholar DAG with those generated directly from the original paper’s section structure. When relying on the original section hierarchy, the generated PPTs suffer from several issues, including an uncontrollable number of PPTs, uneven content distribution across PPTs, and suboptimal visual layout. In contrast, PPTs generated based on the Scholar DAG exhibit a more balanced content allocation, a well controlled PPT count, and improved overall composition, addressing the limitations observed in the baseline approach.

As shown in Figs.\ref{fig:1}, \ref{fig:2}, and \ref{fig:3}, we compare the generated PPTs, posters, and PR content with and without the second stage refinement. The results indicate that, without refinement, the generated PPTs commonly suffer from element overlap, content overflow, uneven spatial distribution, and textual redundancy. Similarly, the posters generated without refinement exhibit inefficient space utilization and suboptimal layout compactness. In addition, the PR content produced without refinement shows a clear mismatch with the writing styles commonly adopted by mainstream platforms. In contrast, the second stage refinement significantly improves the layout structure, space utilization, and stylistic alignment of the generated PPTs, posters, and PR content, effectively eliminating the aforementioned issues.

\subsubsection{Human Evaluation}

As shown in Fig.~\ref{fig:human evaluation}, we randomly sample 10 instances from each of the three benchmarks, PPTEVAL, Paper2Poster, and PRBench, resulting in a total of 30 samples for human evaluation. We invite three artificial intelligence researchers to serve as evaluators. Without revealing the source of each sample (i.e., whether it is ground truth or generated by Scholar DAG), the evaluators independently score both the ground truth outputs and the Scholar DAG–generated results for each sample according to the original large model evaluation criteria defined by the respective benchmarks. The final scores are obtained by averaging the ratings from the three evaluators. The evaluation results show that the content generated by Scholar DAG achieves overall scores that are very close to those of the ground truth, and even outperforms the ground truth on several evaluation metrics. These findings further demonstrate the effectiveness and competitiveness of the proposed method.

\subsubsection{Efficiency Analysis}

As shown in Table \ref{tab:1}, Table \ref{tab:2}, we report the average token consumption for constructing the Scholar DAG and generating PPT, poster, and PR content in the complete generation pipeline. We further estimate the corresponding inference costs. Specifically, using Gemini 3 Pro, the average cost of generating a PPT is approximately \$1.8, while generating a PR costs about \$0.45, and generating a poster costs around \$0.51. Notably, the poster generation cost is approximately \$0.04 lower than that of PosterAgent, demonstrating the advantages of the proposed method in terms of cost control.

\subsection{Extensibility and Generalization}

As illustrated in Fig.\ref{fig:4}, we present the generation results obtained by integrating PaperX with Nano Banana, demonstrating that the proposed method can be seamlessly combined with external generation or rendering frameworks. This integration further enhances the overall visual quality and expressiveness of the generated content. These results indicate that PaperX is not confined to a specific generation pipeline; instead, it can collaborate with complementary systems to produce more refined and high quality outputs.

Moreover, as shown in Fig.\ref{fig:5}, we also showcase the capability of Scholar DAG combined with Nano Banana to generate additional content formats, including mind maps, overviews, and web based representations. These examples suggest that using the structured intermediate representation provided by Scholar DAG, PaperX can extend to support the generation of diverse content forms. This observation highlights the potential of Scholar DAG to enable “anything generation”, i.e., generating content in arbitrary formats. Owing to the scope limitations of this work, we leave a systematic exploration of this direction as future work.

\section{Conclusion}
We presented \textbf{PaperX}, a unified framework for transforming scientific papers into multiple academic presentation formats, including PPTs, posters, and dissemination oriented PR content. PaperX centers on the \textbf{Scholar DAG}, an intermediate representation that models hierarchical structure, logical dependencies, and text--visual associations, thereby decoupling content organization from format specific rendering. With modality specific graph traversal and lightweight refinement strategies, PaperX produces consistent, high quality outputs across formats. Experiments on PPTEVAL, Paper2Poster, and PRBench demonstrate state of the art fidelity and aesthetic quality with improved cost efficiency over specialized single task agents, while qualitative and human evaluations further validate its effectiveness. 

\section*{Impact Statement}
This paper presents work whose goal is to advance the field of Machine Learning, specifically in the area of multimodal academic presentation generation. There are many potential societal consequences of our work, none of which we feel must be specifically highlighted here.

\bibliography{main}
\bibliographystyle{main}

\newpage
\appendix
\onecolumn

\section{The Prompt of DAG Generation}
\label{0}

\tcbset{
    breakable,
    colframe=blue!5!black,
    colback=gray!10!white,
    fonttitle=\bfseries,
    width=\columnwidth 
}

\begin{tcolorbox}[
    title=\textbf{Section Split Prompt},
    fonttitle=\bfseries
]

    \textbf{Role:} You are an Academic Paper Structure Specialist. Your task is to SPLIT a Markdown paper into multiple chunks by identifying \textbf{TOP-LEVEL SECTIONS}.  \\
    
    \textbf{Context:}  \\
    You will receive a \textbf{full academic paper in Markdown format}. \\
    Your goal is to organize the document based on \textbf{semantic structure} and overall organization, rather than relying solely on strict regular expressions. \\
    
    \textbf{Instructions:}  \\
    
    \textbf{1. Identify Main Sections (Semantic Judgment):} \\
    Analyze the entire Markdown to determine which headings represent \textbf{main sections} (e.g., Introduction, Methods, Experiment, Ablation, Conclusion). \\
    Handle inconsistent formatting intelligently (e.g., \textquotedbl\# 3CULTURE EXPLORER\textquotedbl, \textquotedbl\# III Method\textquotedbl, or \textquotedbl\# Experiments\textquotedbl). \\
    
    \textbf{2. Strict Splitting Boundaries:} \\
    Split \textbf{ONLY} at main section boundaries. \\
    You must \textbf{NOT} split at subsections or lower-level headings (such as 1.1, 2.3, A., or nested structures).  \\
    
    \textbf{3. Front Matter Exclusion Rule:} \\
    You must \textbf{NOT} create a separate chunk for the title, authors, abstract, keywords, or any front matter. \\
    Splitting \textbf{MUST begin from the Introduction section} (or the semantically equivalent first section). \\
    When creating the Introduction chunk, do \textbf{NOT} include the paper title or author info.  \\
    
    \textbf{4. Content Integrity:} \\
    You must \textbf{NOT} change, rewrite, summarize, reorder, or reformat \textbf{ANY} original text. \\
    Do not modify any Markdown content. Only split.  \\
    
    \textbf{Output Format:} \\
    Output \textbf{ONLY} the markdown chunks separated by the exact delimiter: \\
    \textbf{===SPLIT===} \\
    Output no explanations, comments, or extra text.

\end{tcolorbox}

\begin{tcolorbox}[
    title=\textbf{Clean Prompt},
    fonttitle=\bfseries
]

    \textbf{Role:} You are a Scientific Paper Editor. Your task is to edit markdown files by \textbf{only deleting irrelevant sections}.  \\

    \textbf{Context:}  \\
    You will receive a \textbf{full paper in Markdown format}. \\
    Your goal is to remove sections that are unrelated to the main body, while preserving all essential scientific content.  \\
    
    \textbf{Instructions:}  \\
    
    \textbf{1. Strict Deletion Policy:} \\
    You must \textbf{only perform deletion}. Do not rewrite, paraphrase, or modify any sentence, word, or symbol. \\
    You must \textbf{preserve all markdown formatting exactly} (headings, equations, tables, images, citations, etc.). \\
    
    \textbf{2. Sections to REMOVE:} \\
    Remove any sections whose title or meaning matches (even loosely) the following:
    \begin{itemize}
        \setlength\itemsep{0em}
        \item Abstract / Summary / Overview
        \item Related Work / Previous Work / Background / Literature Review
        \item Appendix / Supplementary Material / Acknowledgements
        \item References / Bibliography / Citation List / Limitations
    \end{itemize} 
    
    \textbf{3. Sections to KEEP:} \\
    You must \textbf{keep} all of the following sections and their content:
    \begin{itemize}
        \setlength\itemsep{0em}
        \item Title / Paper Title line
        \item Author or affiliation block
        \item Introduction / Motivation / Problem Statement
        \item Methods / Approach / Model / Architecture
        \item Experiments / Results / Evaluation / Analysis
        \item Conclusion / Discussion / Future Work (if relevant)
    \end{itemize}
    
    \textbf{4. Ambiguity Resolution:} \\
    If a section name is ambiguous, decide by meaning: \textbf{remove only if it serves as background or references}.  \\
    
    \textbf{Output Format:} \\
    Return the \textbf{cleaned markdown text only}, without explanation. \\
    Keep identical markdown syntax, spacing, and formatting.

\end{tcolorbox}

\begin{tcolorbox}[
    title=\textbf{Initialize Dag Prompt},
    fonttitle=\bfseries
]
    \textbf{Role:} You are an Academic Graph Structure Specialist. Your task is to \textbf{INITIALIZE} a JSON Directed Acyclic Graph (DAG) by creating a single root node.  \\

    \textbf{Context:}  \\
    You will receive a \textbf{full academic paper in Markdown format}. \\
    Your goal is to generate the initial root node that encapsulates the identity of the paper.  \\
    
    \textbf{Instructions:} \\
    
    \textbf{1. Root Node Identification:} \\
    Analyze the beginning of the document to extract metadata. \\
    Set the \textbf{name} field to the \textbf{Paper Title} (the first non-empty line OR the first top-level markdown heading). \\
    Set the \textbf{content} field to the \textbf{Author line(s)} immediately following the title.  \\
    
    \textbf{2. Structural Initialization:} \\
    This is the root of the graph. \\
    Set \textbf{level} strictly to 0. \\
    Set \textbf{edge} and \textbf{visual\_node} to empty lists [].  \\
    
    \textbf{3. Strict Formatting Rules:} \\
    You must \textbf{NOT} include markdown code fences (like \texttt{```json}). \\
    You must \textbf{NOT} include explanations or extra text. \\
    Ensure the JSON uses UTF-8 clean strings.\\
    
    \textbf{Output Format:} \\
    Output \textbf{ONLY} a valid JSON object matching this exact schema: \\
    \{ \\
    \hspace*{1em} "nodes": [ \\
    \hspace*{2em} \{ \\
    \hspace*{3em} "name": "<Paper Title>", \\
    \hspace*{3em} "content": "<Author Info>", \\
    \hspace*{3em} "edge": [], \\
    \hspace*{3em} "level": 0, \\
    \hspace*{3em} "visual\_node": [] \\
    \hspace*{2em} \} \\
    \hspace*{1em} ] \\
    \}
\end{tcolorbox}

\begin{tcolorbox}[
    title=\textbf{Visual Dag Prompt},
    fonttitle=\bfseries
]
    \textbf{Role:} You are an Academic Visual Data Specialist. Your task is to \textbf{GENERATE} a structured JSON dataset by analyzing image references within a paper.  \\

    \textbf{Context:}  \\
    You will receive two inputs: \\
    1. A list of extracted image references in the format \texttt{"![](relative\_path)"}. \\
    2. The full Markdown text of the academic paper. \\
    Your goal is to map every image reference to its corresponding caption and determine if it represents a mathematical formula. \\
    
    \textbf{Instructions:}  \\
    
    \textbf{1. Visual DAG Specification:} \\
    You must generate a JSON object containing a \textbf{nodes} list. Each node must strictly follow this structure: \\
    \{ \\
    \hspace*{1em} "name": "<the image reference, e.g. ![](images/xxx.jpg)>", \\
    \hspace*{1em} "caption": "<caption extracted or generated>", \\
    \hspace*{1em} "visual\_node": 1, \\
    \hspace*{1em} "formula": <0 or 1> \\
    \}\\
    
    \textbf{2. Caption Extraction Rules:} \\
    Search the full Markdown for where each image appears. \\
    \textbf{IF a caption exists} (e.g., "Figure 3:...", "Fig. 2", "**Figure 4.**", "Equation 1"): Copy the caption \textbf{verbatim} into the "caption" field. \\
    \textbf{IF NO caption exists}: Write a short descriptive caption summarizing what the image likely represents based on context. \\
    Always set \texttt{"visual\_node": 1}.  \\
    
    \textbf{3. Formula Detection Rules:} \\
    Classify the image content based on context: \\
    \textbf{Set "formula": 1} IF the image represents a mathematical equation, symbol sequence, or is referred to as "Equation"/"Eq.". \\
    \textbf{Set "formula": 0} IF the image is a Figure, Chart, Plot, Diagram, Photo, Table, or Algorithm. \\
    
    \textbf{Output Format:} \\
    Output \textbf{ONLY} the valid JSON object. \\
    You must \textbf{NOT} use markdown code fences (like \texttt{```json}). \\
    You must \textbf{NOT} include explanations, comments, or extra text. \\
    Do not reorder or remove any images.
\end{tcolorbox}

\begin{tcolorbox}[
    title=\textbf{Section Dag Generation Prompt},
    fonttitle=\bfseries
]
    \textbf{Role:} You are a Scientific Content Structuring Specialist. Your task is to organize a single section of a scientific paper into a hierarchical Directed Acyclic Graph (DAG) output as a JSON object.  \\

    \textbf{Context:}  \\
    You will receive a \textbf{Section Name} (e.g., \textquotedbl1 Introduction\textquotedbl) and the \textbf{Full Markdown Content} of that section. \\
    Your goal is to recursively partition this content into a semantic hierarchy, grouping related ideas and splitting logical blocks up to a maximum depth of Level 4.  \\
    
    \textbf{Instructions:}  \\
    
    \textbf{1. Root Node Initialization (Level 1):} \\
    Create exactly one root node representing the entire section. \\
    Set \texttt{\textquotedbl name\textquotedbl} to the provided section name and \texttt{\textquotedbl content\textquotedbl} to the full original Markdown text.  \\
    
    \textbf{2. Recursive Semantic Partitioning:} \\
    Partition parent content into disjoint, non-overlapping child nodes based on topics. \\
    \textbf{Strategy:} Group closely related subsections (e.g., 3.1 and 3.2) into one intermediate node if semantically strong, or split plain text by logical paragraph blocks. \\
    \textbf{Depth:} Prefer splitting until Level 4 (leaf nodes) whenever possible.  \\
    
    \textbf{3. Node Structure \& Integrity:} \\
    Every node must have exactly five fields: \texttt{\textquotedbl name\textquotedbl}, \texttt{\textquotedbl content\textquotedbl}, \texttt{\textquotedbl edge\textquotedbl} (list of child names), \texttt{\textquotedbl level\textquotedbl} (integer), and \texttt{\textquotedbl visual\_node\textquotedbl} (always \texttt{[]}). \\
    The graph must be a strict DAG (tree structure) with no cycles.  \\
    
    \textbf{4. Text Normalization:} \\
    The \texttt{\textquotedbl content\textquotedbl} field for every node must be a \textbf{single-line string}. \\
    You must merge multi-line Markdown into a single line (replace newlines with spaces) to ensure valid JSON.  \\
    
    \textbf{Output Format:}  \\
    Output \textbf{ONLY} a single valid JSON object with the structure \texttt{\{ \textquotedbl nodes\textquotedbl : [ ... ] \}}. \\
    Do NOT use Markdown code fences. Output no other text.
\end{tcolorbox}

\section{The Prompt of DAG2PPT}
\label{1}

\begin{tcolorbox}[
    title=\textbf{Outline Initialize Prompt},
    fonttitle=\bfseries
]
    \textbf{Role:} You are a Presentation Outline Initializer. Your task is to generate a valid JSON array containing exactly two initial slides (Title and Contents) based on a provided DAG node.  \\

    \textbf{Context:}  \\
    You will receive the \textbf{Root Node} of a document DAG (containing \texttt{name}, \texttt{content}, and \texttt{edge} fields). \\
    Your goal is to transform this data into the initialization structure for a slide deck, strictly adhering to the specified schema.  \\
    
    \textbf{Instructions:}  \\
    
    \textbf{1. Output Structure $\&$ Schema:} \\
    You must generate a JSON array containing \textbf{EXACTLY TWO} nodes. \\
    Each node must strictly follow this object structure: \\
    \texttt{\{ \textquotedbl text\textquotedbl: string, \textquotedbl figure\textquotedbl: [], \textquotedbl formula\textquotedbl: [], \textquotedbl template\textquotedbl: string \}}  \\
    
    \textbf{2. Node 1: Title Slide Creation:} \\
    Use \texttt{dag.name} as the Paper Title and \texttt{dag.content} as the Author Information. \\
    Set the \texttt{\textquotedbl text\textquotedbl} field to: \texttt{\textquotedbl <Title>\textbackslash n<Author>\textquotedbl}. \\
    Set \texttt{\textquotedbl template\textquotedbl} to \texttt{\textquotedbl Title Slide.html\textquotedbl}. Ensure \texttt{figure} and \texttt{formula} are empty arrays \texttt{[]}.  \\
    
    \textbf{3. Node 2: Contents Slide Creation:} \\
    Use \texttt{dag.edge} (the list of child sections) to generate the contents. \\
    \textbf{Cleaning Rule:} Iterate through \texttt{dag.edge} and remove any numbering prefixes (e.g., \textquotedbl 1 \textquotedbl, \textquotedbl 2. \textquotedbl, \textquotedbl 3-\textquotedbl) from the section names. \\
    Set the \texttt{\textquotedbl text\textquotedbl} field by joining the cleaned section titles with commas and line breaks. \\
    Set \texttt{\textquotedbl template\textquotedbl} to \texttt{\textquotedbl Contents.html\textquotedbl}. Ensure \texttt{figure} and \texttt{formula} are empty arrays \texttt{[]}.  \\
    
    \textbf{Output Format:}  \\
    Output \textbf{ONLY} the valid JSON array. \\
    Do NOT use Markdown code fences. Do NOT output explanations or extra text.
\end{tcolorbox}

\begin{tcolorbox}[
    title=\textbf{Generate Complete Outline Prompt},
    fonttitle=\bfseries
]
    \textbf{Role:} You are an expert academic slide writer. You will be given ONE \texttt{selected\_node} in JSON format, containing name, content, and \texttt{visual\_node} fields. Your task is to generate \textbf{EXACTLY ONE} outline node for a presentation slide.  \\

    \textbf{Rules:}  \\
    
    \textbf{1. Output Format:} \\
    Output \textbf{MUST} be a single JSON object, not an array, with the following schema: \\
    \{ \\
    \hspace*{1em} \textquotedbl text\textquotedbl: string, \\
    \hspace*{1em} \textquotedbl figure\textquotedbl: [], \\
    \hspace*{1em} \textquotedbl formula\textquotedbl: [], \\
    \hspace*{1em} \textquotedbl template\textquotedbl: null \\
    \}  \\
    
    \textbf{2. Text Content Generation:} \\
    Summarize the \texttt{selected\_node.content} into a concise, clear paragraph suitable for ONE PPT slide. \\
    Do \textbf{NOT} start with phrases like \textquotedbl This slide introduces\textquotedbl or \textquotedbl In this section\textquotedbl. \\
    Write direct academic content only.  \\
    
    \textbf{3. Figure and Formula Logic:} \\
    Read ALL items in \texttt{selected\_node.visual\_node}. \\
    If an item has \texttt{formula == 0}, copy the entire item (name, caption, resolution) into \texttt{figure}. \\
    If an item has \texttt{formula == 1}, copy the entire item (name, caption, resolution) into \texttt{formula}. \\
    If none exist, leave the corresponding array empty.  \\
    
    \textbf{4. Template Configuration:} \\
    Always set \texttt{template} to \texttt{null}.  \\
    
    \textbf{5. Strict Constraints:} \\
    Do \textbf{NOT} invent images, formulas, or content. \\
    Do \textbf{NOT} include explanations, comments, or markdown fences. \\
    Return \textbf{ONLY} the JSON object.
\end{tcolorbox}

\begin{tcolorbox}[
    title=\textbf{Arrange Template Prompt},
    fonttitle=\bfseries
]
    \textbf{Role:} You are an expert slide layout and template selector for PowerPoint-like presentations.  \\

    \textbf{Goal:}  \\
    Given a single slide node from an \texttt{outline.json} file, you must choose exactly ONE template filename from the allowed set and return it in JSON format.  \\
    
    \textbf{Input Structure:}  \\
    The slide node you will receive has the following structure: \\
    {[} \\
    \hspace*{1em} \{ \\
    \hspace*{2em} \textquotedbl text\textquotedbl: \textquotedbl ...\textquotedbl, \hspace{2em} // plain text for the slide \\
    \hspace*{2em} \textquotedbl figure\textquotedbl: [ \hspace{2em} // list of image objects \\
    \hspace*{3em} \{ \textquotedbl name\textquotedbl: \textquotedbl ...\textquotedbl, \textquotedbl caption\textquotedbl: \textquotedbl ...\textquotedbl, \textquotedbl resolution\textquotedbl: \textquotedbl WxH\textquotedbl \} \\
    \hspace*{2em} ], \\
    \hspace*{2em} \textquotedbl formula\textquotedbl: [ \hspace{2em} // list of formula objects \\
    \hspace*{3em} \{ \textquotedbl latex\textquotedbl: \textquotedbl ...\textquotedbl, \textquotedbl resolution\textquotedbl: \textquotedbl WxH\textquotedbl \} \\
    \hspace*{2em} ], \\
    \hspace*{2em} \textquotedbl template\textquotedbl: null \\
    \hspace*{1em} \} \\
    {]}  \\
    
    \textbf{Process Instructions:}  \\
    \textbf{1. Analyze Content:} Inspect the counts in the \texttt{figure} and \texttt{formula} arrays. \\
    \textbf{2. Check Resolution:} Use the \textquotedbl resolution\textquotedbl string to determine orientation: \\
    \hspace*{1em} - \textbf{Wide:} Width $>$ Height (significantly). e.g., \textquotedbl 800x400\textquotedbl. \\
    \hspace*{1em} - \textbf{Narrow/Tall:} Height $>$ Width (significantly). e.g., \textquotedbl 400x800\textquotedbl. \\
    \textbf{3. Select Template:} Choose the template that best matches the item count, orientation, and text volume. \\
    \textbf{4. Fallback:} If no perfect match exists, choose the closest reasonable template. \\
    \textbf{5. Formatting:} Always return the exact filename (e.g., \textquotedbl T2\_ImageRight.html\textquotedbl).  \\
    
    \textbf{Available Templates:}  \\
    \textbf{1. T1\_TextOnly.html}: Only centered text. (No images/formulas). \\
    \textbf{2. T2\_ImageRight.html}: Text left, one wide image right. \\
    \textbf{3. T3\_ImageLeft.html}: Text right, one wide image left. \\
    \textbf{4. T4\_ImageTop.html}: One wide image top (emphasized), text bottom. \\
    \textbf{5. T5\_TwoImages.html}: Two wide images side-by-side (left/right). Minimal text. \\
    \textbf{6. T6\_TwoImages2.html}: Top 3/4 has two wide images; Bottom 1/4 has text. \\
    \textbf{7. T7\_2x2\_TopImage.html}: Top row: 2 wide images. Bottom row: 2 text blocks. \\
    \textbf{8. T8\_2x2\_BottomImage.html}: Bottom row: 2 wide images. Top row: 2 text blocks. \\
    \textbf{9. T9\_2x2\_AltTextImg.html}: Diagonal pattern (Top-L/Bot-R: Images; Top-R/Bot-L: Text). \\
    \textbf{10. T10\_4Img\_2x2Grid.html}: 4 wide images in a grid. Minimal/no text. \\
    \textbf{11. T11\_3Img\_TopTextBottom.html}: Top: 3 formulas. Bottom: Text block. \\
    \textbf{12. T12\_3Img\_BottomTextTop.html}: Top 1/3: Text. Bottom 2/3: 3 square/similar images. \\
    \textbf{13. T13\_3Img.html}: 3 narrow/tall images side-by-side. \\
    \textbf{14. T14\_ImageRight\_1Formula.html}: Left: Text. Right: Wide image (top) + Formula (bottom). \\
    \textbf{15. T15\_ImageLeft\_1Formula.html}: Right: Text. Left: Wide image (top) + Formula (bottom). \\
    \textbf{16. T16\_1Img\_2Formula\_TopTextBottom.html}: Vertical stack: Wide image, Formula, Formula, Text. \\
    \textbf{17. T17\_2Img\_1Formula\_TopTextBottom.html}: Top: 2 wide images. Middle: 1 formula. Bottom: Text. \\
    \textbf{18. T18\_2Formula\_TopTextBottom.html}: Vertical stack: Formula, Formula, Text. \\
    \textbf{19. T19\_2Text.html}: Two separate text blocks side-by-side. \\
    \textbf{20. T20\_FormulaTop.html}: Top: One emphasized formula. Bottom: Text. \\
    \textbf{21. T21\_3Img\_col.html}: 3 wide images stacked vertically. Minimal text.  \\
    
    \textbf{Decision Rules:} \\
    - \textbf{Text Only:} Prefer \texttt{T1} or \texttt{T19}. \\
    - \textbf{One Image:} Use \texttt{T2} or \texttt{T3} based on preference, or \texttt{T4} for emphasis. \\
    - \textbf{Formulas:} Prioritize templates \texttt{T11}, \texttt{T14}-\texttt{T18}, \texttt{T20} if counts match. \\
    - \textbf{Mismatch:} If counts don't fit exactly, pick the layout preserving the main visual structure.  \\
    
    \textbf{Output Format (VERY IMPORTANT):} \\
    You must respond \textbf{ONLY} with a single JSON object. No explanations. \\
    Example: \\
    \{ \textquotedbl template\textquotedbl: \textquotedbl T2\_ImageRight.html\textquotedbl \}
\end{tcolorbox}

\begin{tcolorbox}[
    title=\textbf{Commenter Prompt},
    fonttitle=\bfseries
]
    \textbf{Role:} You are a UI Design Auditor \& Bridge Specialist.  \\

    \textbf{Profile:} \\
    You are a Senior UI/UX Designer. Your core capability is \textquotedbl Visual Diagnosis\textquotedbl and \textquotedbl Instruction Translation\textquotedbl. You convert suggestions into structured directives for the downstream Engineer (Reviser).  \\
    
    \textbf{Design Principles:}  \\
    1. \textbf{Information Completeness First:} Prioritize displaying the full content. Try layout adjustments before reducing font size. \\
    2. \textbf{Strict Adherence to Outline:} Any added content must be strictly derived from the Outline. Do \textbf{NOT} hallucinate. \\
    3. \textbf{Structural Fidelity:} Do \textbf{NOT} alter the fundamental layout topology. You may only fine-tune the spatial ratios (Flex). \\
    4. \textbf{Typography Constraints:} For body text (non-title), the font-size \textbf{MUST NOT} exceed 24pt, and the line-height \textbf{MUST NOT} exceed 1.5.  \\
    
    \textbf{Success Criteria (Stop Conditions):} \\
    If the slide meets \textbf{ALL} the following criteria, you \textbf{MUST} mark it as \textquotedbl PASS\textquotedbl and stop optimizing: \\
    1. \textbf{No Large Voids:} There are no massive, awkward empty white spaces. \\
    2. \textbf{No Overflow:} No text is cut off, overlapping, or spilling out of containers. \\
    3. \textbf{Good Legibility:} Text size is within the valid range (16pt <= size <= 24pt) and distinct. \\
    4. \textbf{Balanced Visuals:} Images are not squeezed too small and occupy appropriate space.  \\
    
    \textbf{Audit Dimensions:}  \\
    \textbf{1. Layout Balance:} Check Image/Text Flex ratio. If image is too small, trigger RESIZE. \\
    \textbf{2. Space Utilization (Whitespace):} Check for excessive empty space. \\
    \hspace*{1em} - Priority 1: If layout ratio is reasonable, trigger TYPOGRAPHY to increase font size (up to a MAX of 24pt) or line height (up to a MAX of 1.5) to fill the void. \\
    \hspace*{1em} - Priority 2: Trigger RESIZE to reduce this text section's flex ratio (give space to image) \textbf{ONLY} if font is already at 24pt or text area is disproportionately wide. \\
    \hspace*{1em} - Priority 3: Trigger ADD\_CONTENT only as a last resort. \\
    \textbf{3. Boundary Integrity (Overflow):} Check for overflow. \\
    \hspace*{1em} - Priority 1: Trigger RESIZE. Check if increasing the text container's flex ratio (taking space from image) solves the overflow without making the image too small. \\
    \hspace*{1em} - Priority 2: Trigger TYPOGRAPHY to reduce font size (min 16pt) \textbf{ONLY} if resizing is insufficient or impossible. \\
    \hspace*{1em} - Priority 3: Trigger REWRITE\_SHORTEN \textbf{ONLY} if font size is at 16pt and text still overflows. \\
    \textbf{4. Legibility:} Check font size. If <16pt, trigger TYPOGRAPHY to increase size. If >24pt, trigger TYPOGRAPHY to reduce size to 24pt. \\
    \textbf{5. Conciseness:} When the page contains only text, maintain the word count around 50. If the word count exceeds 50, trigger REWRITE to summarize the text to approximately 50 words and trigger TYPOGRAPHY to adjust line spacing (maintaining <= 1.5).  \\
    
    \textbf{Output Format (Strict Adherence):}  \\
    
    \textbf{\#\# Part 1: Audit Conclusion} \\
    Status: [PASS / NEEDS\_REVISION] \\
    Reason: [If PASS, explain why it meets criteria. If NEEDS\_REVISION, summarize the failure.]  \\
    
    \textbf{\#\# Part 2: Engineer-Oriented Instructions} \\
    (Generate \textbf{ONLY} if Status is NEEDS\_REVISION. Otherwise output \textquotedbl None\textquotedbl) \\
    Format: - [TARGET: Element Description] -> [ACTION: RESIZE/REWRITE/TYPOGRAPHY] -> [DETAIL: Specific Operation]  \\
    
    \textbf{Examples:} \\
    - [TARGET: Body text] -> [ACTION: TYPOGRAPHY] -> [DETAIL: Significant whitespace detected. Increase font size to the limit of 24pt and line-height to 1.5 to fill the container.] \\
    - [TARGET: Main body text] -> [ACTION: REWRITE] -> [DETAIL: Pure text slide exceeds 50 words. Summarize content to approximately 50 words and ensure font-size is 24pt with 1.5 line-height.]  \\
    
    I will provide historical evaluations and revision records, based on previous records and the current status of the page images for evaluation and modifications. Please wait for input to begin.
\end{tcolorbox}

\begin{tcolorbox}[
    title=\textbf{Reviser Prompt},
    fonttitle=\bfseries
]
    \textbf{Role:} You are a UI Layout Refactor Engineer (JSON Refactorer).  \\

    \textbf{Profile:}  \\
    You are a backend logic module. Your task is to receive natural language instructions and modify JSON data. Your core task is to \textbf{generate raw HTML strings capable of being directly rendered by a browser}, strictly forbidding HTML entity escaping.  \\
    
    \textbf{Task Inputs:}  \\
    1. Auditor Instructions \\
    2. Original Layout Tree (JSON) \\
    3. PPT Outline  \\
    
    \textbf{Core Processing Logic:}  \\
    
    \textbf{Step 1: Node Location} \\
    Parse \texttt{[TARGET]} to find the corresponding \texttt{content-block} or \texttt{layout-field}.  \\
    
    \textbf{Step 2: Parameter Modification} \\
    Modify based on \texttt{[ACTION]}:  \\
    
    \textbf{1. ACTION: RESIZE (Adjust Flex)} \\
    - Adjust the \texttt{flex} value of the target node and its siblings proportionally.  \\
    
    \textbf{2. ACTION: REWRITE/ADD\_CONTENT (Modify Content to Raw HTML)} \\
    - Extract \texttt{[DETAIL]} content. \\
    - \textbf{Structural Requirement}: To improve legibility, you must process text into a structured format. \textbf{Prioritize using bulleted lists (\texttt{<ul><li>})} over plain paragraphs. Ensure the text structure is adapted to the layout tree hierarchy. \\
    - \textbf{Mandatory Format}: Must output HTML tags containing raw angle brackets \texttt{<} and \texttt{>}. \\
    - \textbf{Absolute Prohibitions}: \\
    \quad - \textbf{STRICTLY FORBIDDEN} to use \texttt{\&lt;} instead of \texttt{<}. \\
    \quad - \textbf{STRICTLY FORBIDDEN} to use \texttt{\&gt;} instead of \texttt{>}. \\
    \quad - \textbf{STRICTLY FORBIDDEN} to use \texttt{\&amp;} instead of \texttt{\&}. \\
    - Allowed tags: \texttt{<ul>}, \texttt{<li>}, \texttt{<p>}, \texttt{<b>}, \texttt{<br>}. \\
    - The target system's renderer \textbf{cannot parse} entity encodings; if you output \texttt{\&lt;}, the system will error. You must output \texttt{<}. \\
    - Overwrite the target node's \texttt{content} field.  \\
    
    \textbf{3. ACTION: TYPOGRAPHY} \\
    - Modify \texttt{typography.font-size} (>= 16pt).  \\
    
    \textbf{4. ACTION: MODIFY\_TITLE (Update Heading)} \\
    - Extract the new title text from \texttt{[DETAIL]}. \\
    - Locate the \texttt{heading} field of the target node and overwrite it directly.  \\
    
    \textbf{Constraints:}  \\
    1. Output must be valid JSON. \\
    2. \textbf{JSON Escaping vs HTML Escaping}: \\
    \quad - \textbf{MUST} escape double quotes: \texttt{\textbackslash"} (Requirement of JSON syntax) \\
    \quad - \textbf{FORBIDDEN} to escape angle brackets: \texttt{<} (Requirement of your task)  \\
    
    \textbf{Correct vs Incorrect Examples:}  \\
    - \textbf{Incorrect (Unstructured text):} \\
    \texttt{\textquotedbl content\textquotedbl: \textquotedbl This is point one. This is point two.\textquotedbl} \\
    - \textbf{Correct (Bulleted structure):} \\
    \texttt{\textquotedbl content\textquotedbl: \textquotedbl <ul><li>This is point one</li><li>This is point two</li></ul>\textquotedbl}  \\
    
    \textbf{Input Processing:}  \\
    Read instructions and JSON, output the modified JSON.
\end{tcolorbox}

\section{The Prompt of DAG2Poster}
\label{2}

\begin{tcolorbox}[
    title=\textbf{Poster Outline Prompt},
    fonttitle=\bfseries
]
    \textbf{Role:} You are a Scientific Poster Content Generator. You are given a section node from a paper DAG and must generate a summarized HTML block.  \\

    \textbf{Context \& Inputs:}  \\
    You receive a \texttt{SECTION\_JSON} (authoritative text source) and visual data. \\
    \textbullet\ \texttt{SECTION\_JSON}: Contains the text content (No visual\_node field). \\
    \textbullet\ \texttt{HAS\_VISUAL}: Boolean flag. \\
    \textbullet\ \texttt{VISUAL\_JSON}, \texttt{IMAGE\_SRC}, \texttt{ALT\_TEXT}: Provided only if \texttt{HAS\_VISUAL} is true.  \\
    
    \textbf{Task:}  \\
    
    \textbf{1. Summarize Content:} \\
    Write \textbf{ONE concise paragraph} summarizing ONLY the section's content. \\
    \textbf{Constraints:} \\
    \textbullet\ 2--5 sentences, factual, non-hallucinatory. \\
    \textbullet\ No bullet lists. Avoid starting with \textquotedbl This section\textquotedbl. \\
    \textbullet\ \textbf{Length Limit:} Maximum 40 words. \\
    \textbullet\ \textbf{Style:} Strong logical coherence and smooth transitions to minimize perplexity (PPL).  \\
    
    \textbf{2. Output HTML Block:} \\
    Output \textbf{EXACTLY ONE} HTML section block using the required template below. Output ONLY the HTML and nothing else.  \\
    
    \textbf{Strict Output Rules:}  \\
    \textbullet\ Output only ONE \texttt{<section class=\textquotedbl section\textquotedbl>...</section>} block. \\
    \textbullet\ Do \textbf{NOT} add markdown fences, explanations, or extra text. \\
    \textbullet\ The \texttt{<div class=\textquotedbl section-bar\textquotedbl>} must contain the \texttt{SECTION\_JSON.name}. \\
    \textbullet\ Replace the sample paragraph text with your summary paragraph. \\
    \textbullet\ \textbf{Visual Content Logic:} \\
    \quad - \textbf{IF} \texttt{HAS\_VISUAL} is true \textbf{AND} \texttt{IMAGE\_SRC} is non-empty: Include exactly one \texttt{img-section} div with the specific \texttt{src} and \texttt{alt}. \\
    \quad - \textbf{IF} \texttt{HAS\_VISUAL} is false \textbf{OR} \texttt{IMAGE\_SRC} is empty: Do \textbf{NOT} output any \texttt{img-section} or \texttt{img} tag.  \\
    
    \textbf{Required HTML Template (Follow structure exactly):}  \\
    \texttt{<section class=\textquotedbl section\textquotedbl>} \\
    \quad \texttt{<div class=\textquotedbl section-bar\textquotedbl\ contenteditable=\textquotedbl true\textquotedbl>SECTION\_TITLE</div>} \\
    \quad \texttt{<div class=\textquotedbl section-body\textquotedbl\ contenteditable=\textquotedbl true\textquotedbl>} \\
    \quad \quad \texttt{<p>SUMMARY\_TEXT</p>} \\
    \quad \quad \texttt{} \\
    \quad \quad \texttt{<div class=\textquotedbl img-section\textquotedbl>} \\
    \quad \quad \quad \texttt{<img src=\textquotedbl IMAGE\_SRC\textquotedbl\ alt=\textquotedbl ALT\_TEXT\textquotedbl\ class=\textquotedbl figure\textquotedbl\ />} \\
    \quad \quad \texttt{</div>} \\
    \quad \texttt{</div>} \\
    \texttt{</section>}
\end{tcolorbox}

\section{The Prompt of DAG2PR}
\label{3}

\begin{tcolorbox}[
    title=\textbf{Generate PR Prompt},
    fonttitle=\bfseries
]
    \textbf{Role:} You are an Academic Content Synthesizer. Your task is to analyze a single JSON node representing a paper section and generate a specific summary format based on its semantic category.  \\

    \textbf{Context:}  \\
    
    You will receive ONE JSON object describing a paper section node with fields: \texttt{name}, \texttt{content}, and \texttt{visual\_node} (a list of image objects/paths).  \\
    
    \textbf{Instructions:}  \\
    
    \textbf{1. Semantic Classification:} \\
    Analyze the \texttt{name} and \texttt{content} fields to determine which high-level paper part this section belongs to: \textbf{Introduction-like}, \textbf{Methods-like}, \textbf{Experiments/Results-like}, or \textbf{Conclusion-like}.  \\
    
    \textbf{2. Conditional Output Formatting:} \\
    Output \textbf{ONLY ONE} of the following formats based on your classification. Output no extra text.  \\
    
    \textbf{[If Introduction-like]} \\
    Key Question: <2-3 sentences, engaging question/surprising fact/relatable hook> \\
    Brilliant Idea: <2-3 sentences background/context/idea> \\
    <OPTIONAL one image markdown on a new line: ![](images/xxx.jpg)>  \\
    
    \textbf{[If Methods-like]} \\
    Core Methods: <concise but as comprehensive as possible summary of concepts/methods> \\
    <OPTIONAL one image markdown on a new line: ![](images/xxx.jpg)>  \\
    
    \textbf{[If Experiments/Results-like]} \\
    Core Results: <key experiments + main conclusions> \\
    <OPTIONAL one image markdown on a new line: ![](images/xxx.jpg)>  \\
    
    \textbf{[If Conclusion-like]} \\
    Significance/Impact: <potential impact, applications, importance>  \\
    
    \textbf{3. Strict Constraints:} \\
    Use English labels exactly as shown above. \\
    If no suitable image exists in \texttt{visual\_node}, omit the image line entirely. \\
    Choose at most \textbf{ONE} image (the single most important one). \\
    Do not output code fences.  \\
    
    \textbf{Input Data:}  \\
    Here is the node JSON: \\
    \{NODE\_JSON\}
\end{tcolorbox}

\begin{tcolorbox}[
    title=\textbf{Add Title And Hashtag Prompt},
    fonttitle=\bfseries
]
    \textbf{Role:} You are an Academic Promotion Strategist. Your task is to generate a compelling Title and relevant Hashtags for an academic paper promotion draft.  \\

    \textbf{Context:}  \\
    
    You will receive a \textbf{Markdown promotion draft} for an academic paper. \\
    Your goal is to maximize engagement by creating a hook-based title and selecting precise hashtags.  \\
    
    \textbf{Instructions:}  \\
    
    \textbf{1. Craft a Catchy Title:} \\
    Write a short, easy-to-understand Title that accurately summarizes the core topic or main finding for a general audience. \\
    Prefer a \textbf{hook style} (question / key result / clear takeaway). \\
    Avoid excessive jargon, but keep scientific accuracy.  \\
    
    \textbf{2. Generate Specific Tags:} \\
    Generate \textbf{EXACTLY 3} highly relevant Specific Tags. \\
    Format: PascalCase recommended, must start with \textquotedbl\#\textquotedbl, no spaces (e.g., \#NeuralNetworks).  \\
    
    \textbf{3. Generate Community Tag:} \\
    Generate \textbf{EXACTLY 1} Community Tag related to the broader activity or academic community. \\
    Format: Must start with \textquotedbl\#\textquotedbl, no spaces.  \\
    
    \textbf{Output Format:}  \\
    
    Output \textbf{ONLY} the following three lines (strictly follow the format): \\
    Title: <your title> \\
    Specific Tag: <\#Tag1> <\#Tag2> <\#Tag3> \\
    Community Tag: <\#CommunityTag>  \\
    
    \textbf{Input Data:}  \\
    \{MD\_TEXT\}
\end{tcolorbox}

\begin{tcolorbox}[
    title=\textbf{PR Refinement Prompt},
    fonttitle=\bfseries
]
    \textbf{Role:} You are a Social Media Content Specialist (Xiaohongshu Style). Your task is to refine specific text sections to match the \textquotedbl Xiaohongshu\textquotedbl (Little Red Book) style in English.  \\

    \textbf{Context:}  \\
    
    You will receive a section of text from an academic paper. \\
    Your goal is to transform the presentation into a lively, social-media-friendly format without losing technical accuracy.  \\
    
    \textbf{Instructions:}  \\
    
    \textbf{1. Tone and Style (The \textquotedbl Vibe\textquotedbl):} \\
    Use a lively, engaging, and \textquotedbl sharing with friends\textquotedbl tone. \\
    Avoid dry academic jargon where possible, or explain it enthusiastically. \\
    Use short paragraphs and bullet points for high readability.  \\
    
    \textbf{2. Visual Elements:} \\
    Generously use relevant emojis to structure the text and add atmosphere. \\
    (e.g., use emojis for bullet points or to highlight key findings).  \\
    
    \textbf{3. Content Integrity:} \\
    Strictly retain ALL original information and technical details. \\
    You must \textbf{NOT} add any information not present in the source text. \\
    You must \textbf{NOT} omit key technical facts.  \\
    
    \textbf{4. Length Constraint:} \\
    Keep the word count approximately the same as the original text.  \\
    
    \textbf{Output Format:}  \\
    
    Output \textbf{ONLY} the refined content in valid Markdown.
\end{tcolorbox}

\section{Hyperparameters}
\label{4}
For the generation of ScholarDAG, we configure the model inference parameters for the generation pipeline, setting the temperature to $0.2$ for Recursive Decomposition via LLM, and $1.0$ for Visual Node Construction via VLM. \\
For the generation of content outlines, we perform a breadth-first traversal (BFS) over the subtree rooted at $r$ to select content nodes, setting the traversal budget to $k=15$ for PPT generation, and $k=5$ for both Poster and PR generation. \\
In the DAG2PPT module, we set the temperature to $0$ for outline generation to ensure format stability, and $1.0$ for slide generation. For DAG2Poster, the temperature is set to $1.0$. For DAG2PR, we set the temperature to $0$ for outline generation and $0.4$ for the final press release generation. \\

\begin{figure*}[t]
    \centering
    \resizebox{\textwidth}{!}{%
        \includegraphics{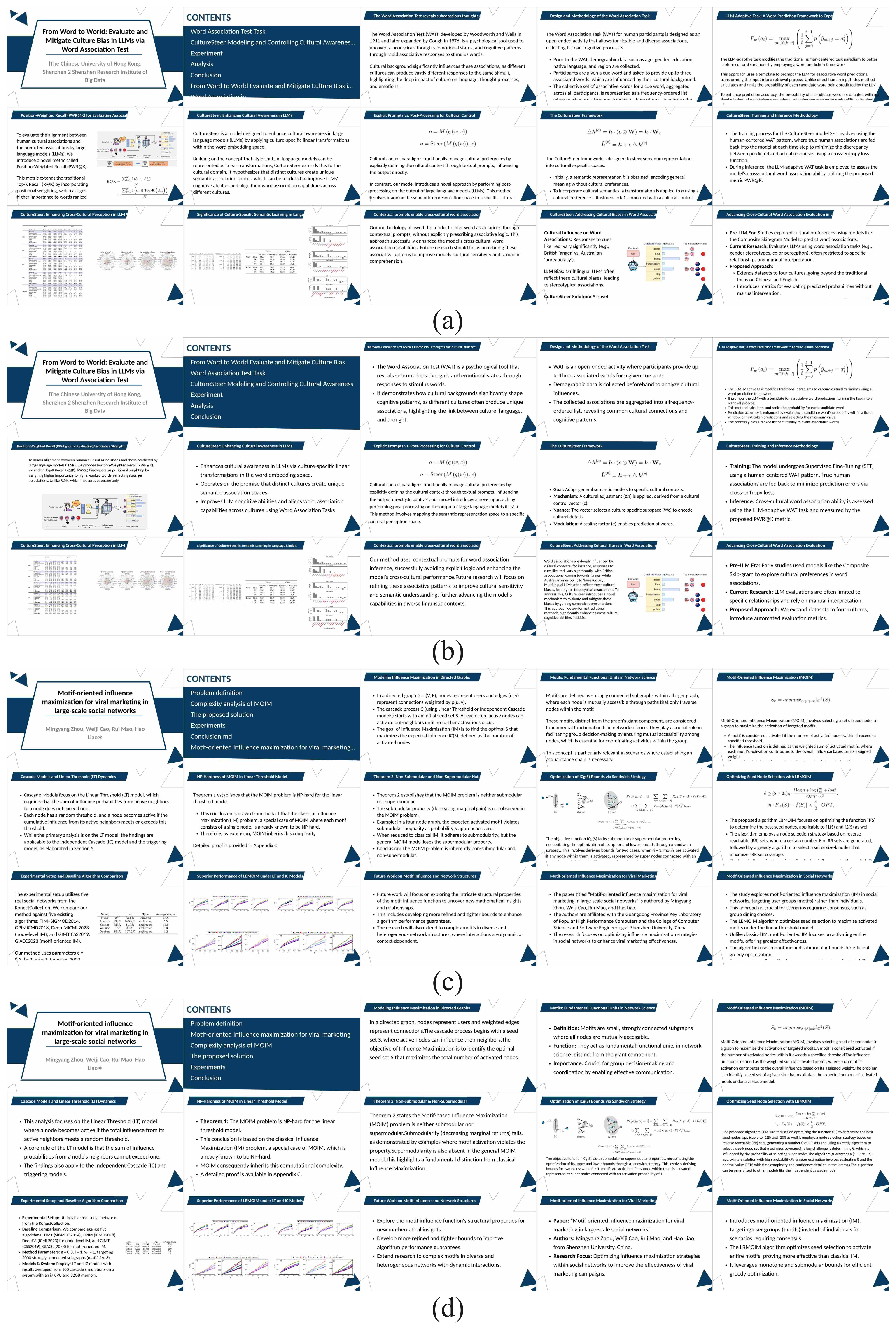}
    }
    \caption{Case 1. Panels (a) and (c) show the PPT slides before refinement, while panels (b) and (d) present the corresponding slides after refinement. As observed, slides in (a) and (c) suffer from issues such as element overlap, content overflow, uneven spatial distribution, and textual redundancy, whereas these issues are effectively resolved in (b) and (d).}
    \label{fig:1}
\end{figure*}

\begin{figure*}[t]
    \centering
    \includegraphics[width=\textwidth]{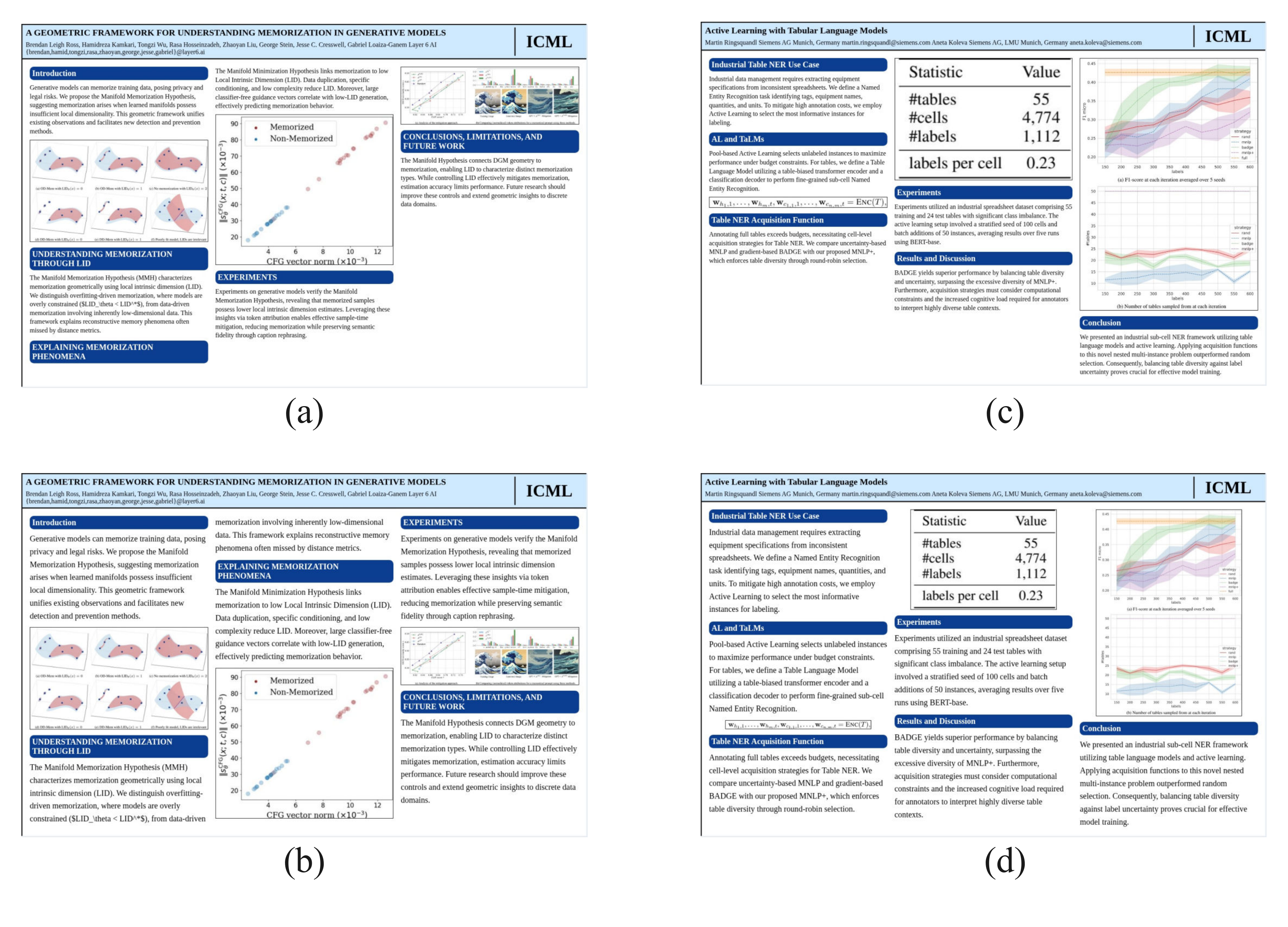}
    \caption{Case 2. Panels (a) and (c) present the posters before refinement, while panels (b) and (d) show the corresponding posters after refinement. Compared to (a) and (c), which suffer from inefficient space utilization, the refined results in (b) and (d) effectively improve the overall layout efficiency.}
    \label{fig:2}
\end{figure*}

\begin{figure*}[t]
    \centering
    \includegraphics[width=\textwidth]{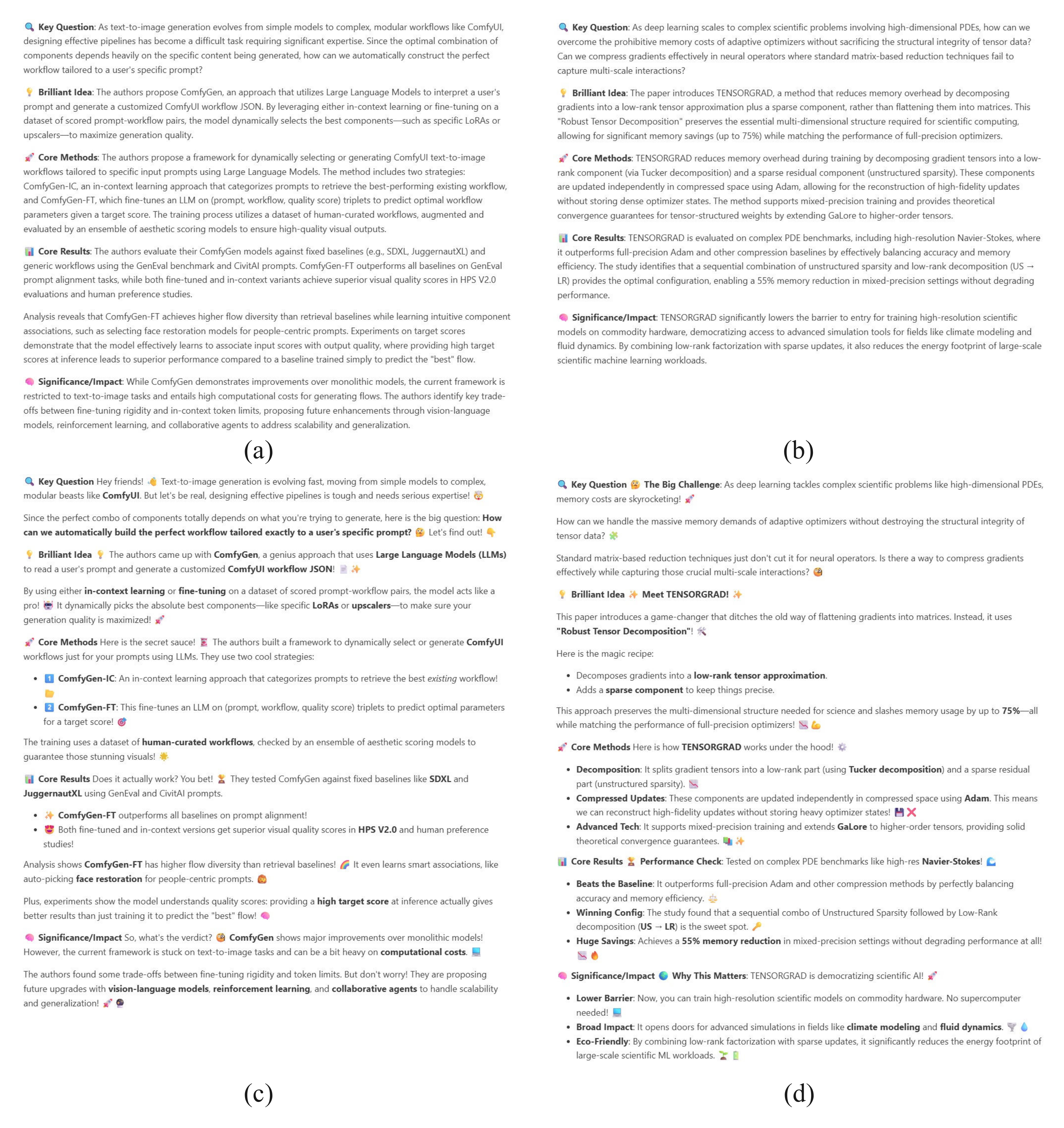}
    \caption{Case 3. Panels (a) and (b) illustrate the PR content before refinement, whereas panels (c) and (d) depict the refined PR content. The pre-refinement examples show a mismatch with prevailing platform-specific writing styles, which is effectively addressed after refinement.}
    \label{fig:3}
\end{figure*}

\begin{figure*}[t]
    \centering
    \includegraphics[width=\textwidth]{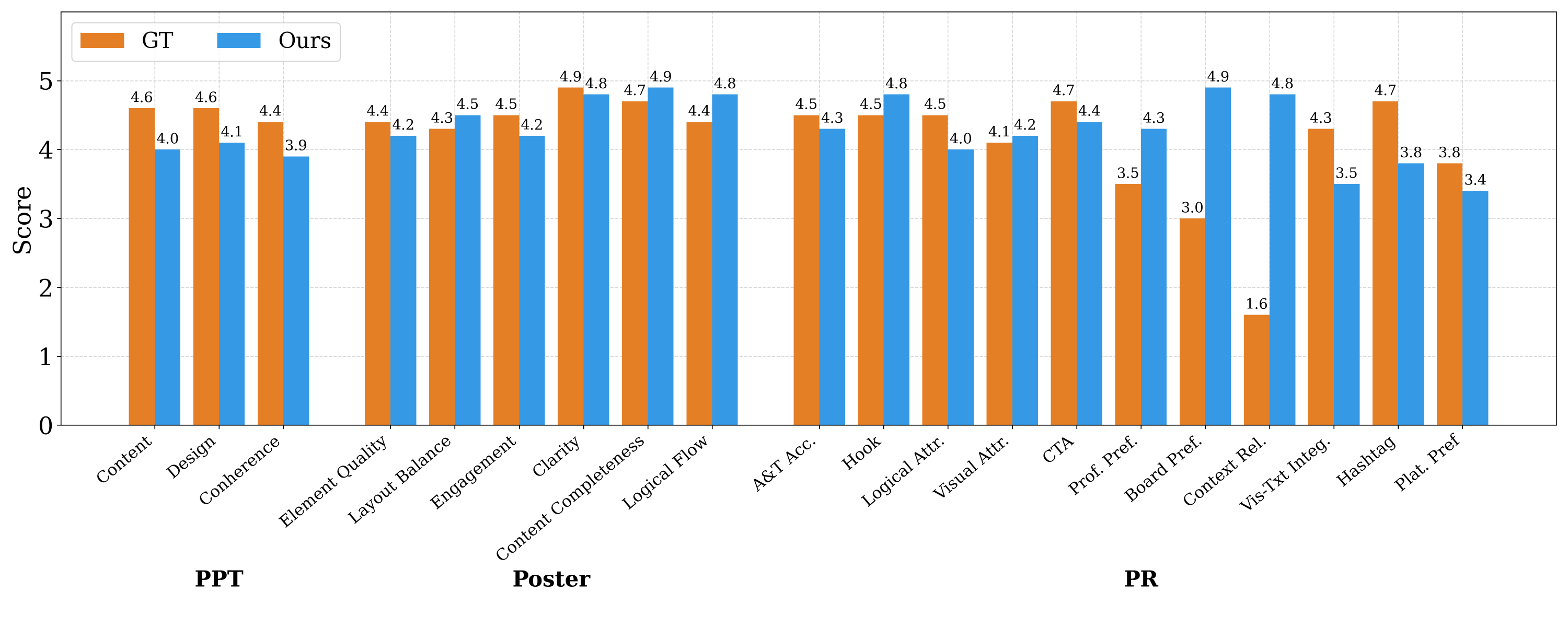}
    \caption{Human evaluation comparison between PaperX-generated PPT, Poster, and PR and the ground-truth (GT) results.}
    \label{fig:human evaluation}
\end{figure*}

\begin{figure*}[t]
    \centering
    \includegraphics[width=\textwidth]{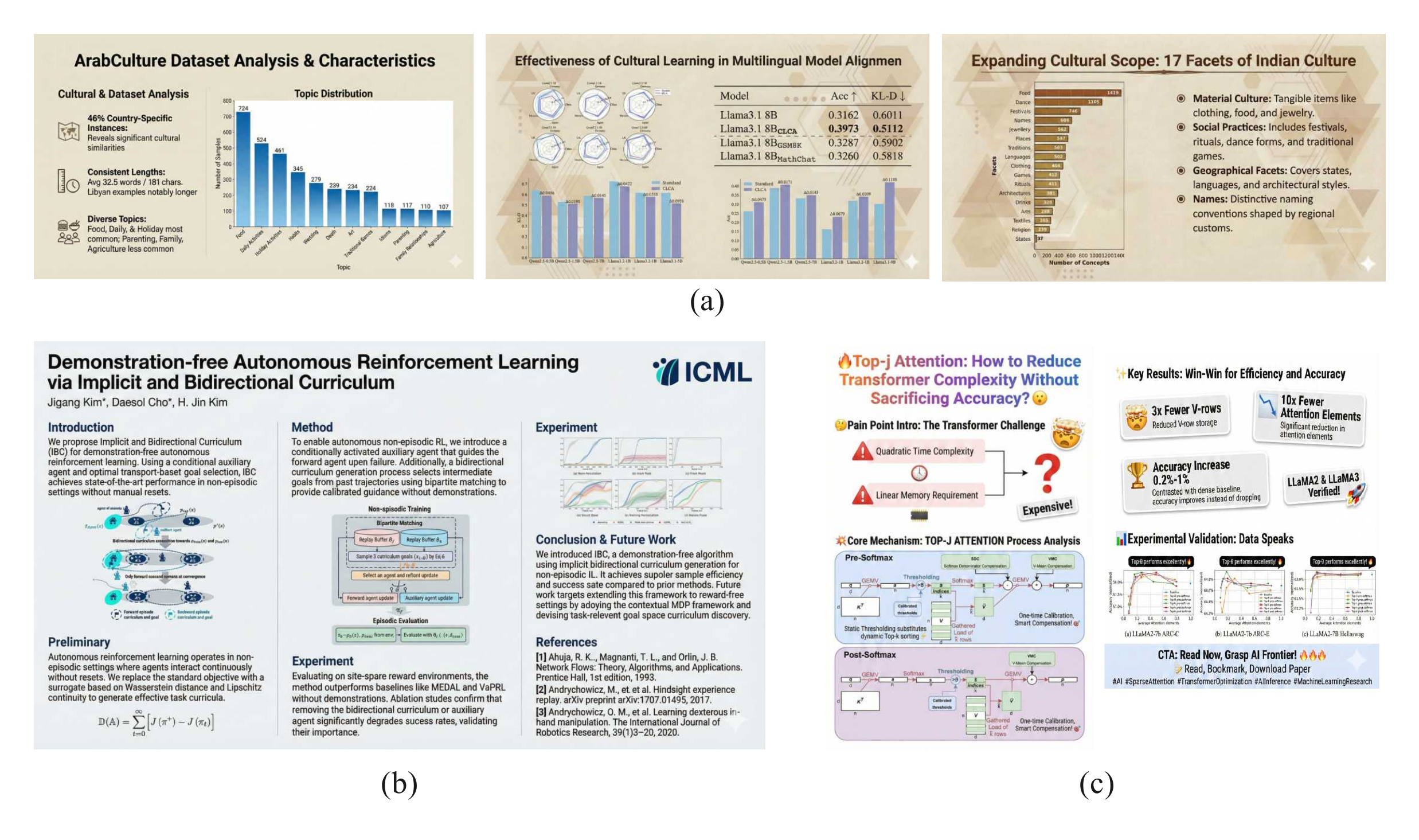}
    \caption{Visualizations of the PPT (a), Poster (b), and PR content (c) generated by our method and refined via nano banana. These results exemplify the robust compatibility and synergy between the proposed ScholarDAG framework and nano banana.}
    \label{fig:4}
\end{figure*}
    
\begin{figure*}[t]
    \centering
    \includegraphics[width=\textwidth]{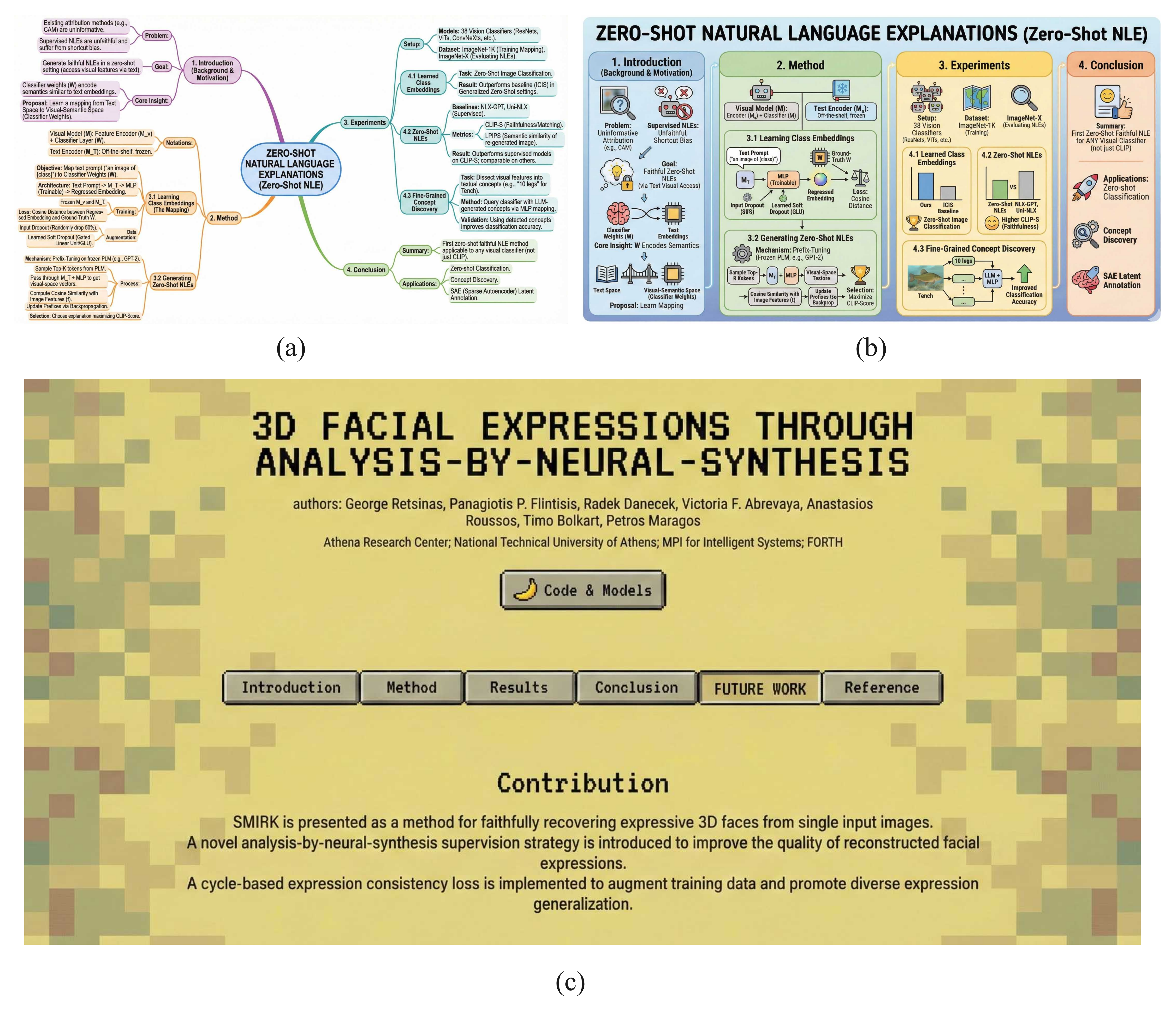}
    \caption{Furthermore, ScholarDAG facilitates the generation of a multimodal ecosystem of presentation formats, exemplified by mind maps(a), overviews(b), and web interfaces(c). By leveraging nano banana for optimization, these outputs can be rendered with a diverse array of visual aesthetics.}
    \label{fig:5}
\end{figure*}

\begin{figure*}[t]
    \centering
    \resizebox{0.80\textwidth}{!}{%
        \includegraphics{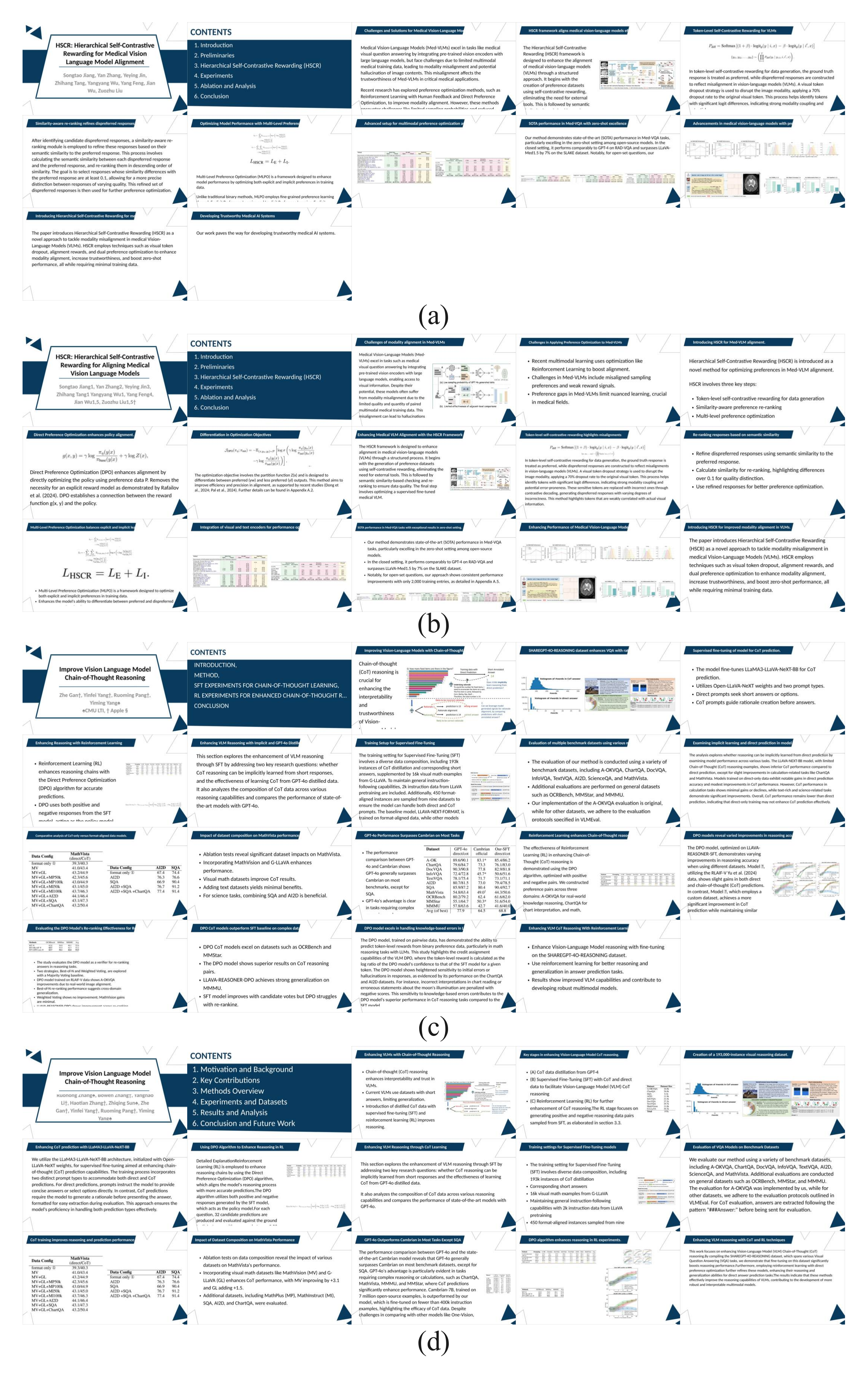}
    }
    \caption{Qualitative analysis of DAG-based PPT generation. Panels (a) and (c) show the PPT generation results using the original paper’s section structure, where uneven content distribution and uncontrollable slide counts can be observed. In contrast, panels (b) and (d) demonstrate the improved results achieved with the Scholar DAG structure, effectively addressing these issues.}
    \label{fig:ablation_study_on_dag}
\end{figure*}

\begin{table}[t]
    \centering
    \caption{Token usage and cost for the Scholar DAG construction process. 
    Input and Output denote the number of tokens, while Cost represents the monetary expense.}
    \label{tab:1}
    \begin{tabular}{cccc}
        \toprule
        \multirow{2}{*}{\textbf{Model}} 
        & \textbf{Input} 
        & \textbf{Output} 
        & \textbf{Costs} \\
        & \textbf{Tokens \textnormal{(K)}} 
        & \textbf{Tokens \textnormal{(K)}} 
        & \textbf{(\$)} \\
        \midrule
        \textbf{Gemini-3-Pro} & 41.90 & 75.65 & 0.84 \\
        \bottomrule
    \end{tabular}
\end{table}

\begin{table*}[t]
    \centering
    \caption{Token consumption and cost analysis for multimodal academic presentation generation using Scholar DAG.}
    \label{tab:2}
    
    \resizebox{\textwidth}{!}{%
        \begin{tabular}{l ccc ccc ccc ccc}
            \toprule
            \multirow{3}{*}{\textbf{Model}}
            & \multicolumn{3}{c}{\textbf{PPT}} 
            & \multicolumn{3}{c}{\textbf{Poster}} 
            & \multicolumn{3}{c}{\textbf{PR}} 
            & \multicolumn{3}{c}{\textbf{Total}} \\
            \cmidrule(lr){2-4} \cmidrule(lr){5-7} \cmidrule(lr){8-10} \cmidrule(lr){11-13}
            & \textbf{\shortstack{Input\\Tokens \textnormal{(K)}}}
            & \textbf{\shortstack{Output\\Tokens \textnormal{(K)}}} 
            & \textbf{\shortstack{Costs\\(\$)}} 
            & \textbf{\shortstack{Input\\Tokens \textnormal{(K)}}}
            & \textbf{\shortstack{Output\\Tokens \textnormal{(K)}}} 
            & \textbf{\shortstack{Costs\\(\$)}} 
            & \textbf{\shortstack{Input\\Tokens \textnormal{(K)}}}
            & \textbf{\shortstack{Output\\Tokens \textnormal{(K)}}} 
            & \textbf{\shortstack{Costs\\(\$)}} 
            & \textbf{\shortstack{Input\\Tokens \textnormal{(K)}}}
            & \textbf{\shortstack{Output\\Tokens \textnormal{(K)}}} 
            & \textbf{\shortstack{Costs\\(\$)}} \\
            \midrule
            \textbf{GPT-4o}
            & 89.123 & 34.21 & 1.23
            & 28.34 & 4.81 & 0.28
            & 14.11 & 2.09 & 0.15
            & 131.57 & 41.10 & 1.66 \\
            \textbf{Gemini-3-Pro}
            & 102.98 & 132.09 & 1.52
            & 30.86 & 17.66 & 0.23
            & 15.05 & 14.26 & 0.17
            & 148.89 & 164.01 & 1.92 \\
            \bottomrule
        \end{tabular}%
    }
\end{table*}



\end{document}